\documentclass[twocolumn,rmp,aps,nofootinbib,amsmath,amssymb,floatfix]{revtex4}

\usepackage{graphics}
\usepackage{epsfig}
\usepackage{longtable}
\usepackage{bm}
\usepackage{color}
\definecolor{Blue}{rgb}{0.3,0.3,0.9}
\definecolor{Red}{rgb}{1,0.1,0.1}
\usepackage[colorlinks,
            linkcolor=black,
            anchorcolor=blue,
            citecolor=blue
            ]{hyperref}

\usepackage{booktabs}

\usepackage{amsmath,mathtools}

\usepackage[normalem]{ulem}

\definecolor{mblue}{rgb}{0,0.35,0.75}

\definecolor{ared}{rgb}{0.7,0.0,0.0}
\newcommand{\addAlexey}[1]{\textcolor{ared}{#1}}

\definecolor{MAred}{rgb}{0.6,0,0}

\definecolor{green}{rgb}{0.0,0.7,0.0}

\begin{document}
\title{Excitons in atomically thin transition metal dichalcogenides}

\author{Gang Wang}
\affiliation{Universit\'e de
Toulouse, INSA-CNRS-UPS, LPCNO, 135 Av. Rangueil, 31077 Toulouse,
France *}

\author{Alexey Chernikov}
\affiliation{Department of Physics, University of Regensburg, D-93040 Regensburg, Germany}

\author{Mikhail M. Glazov}
\affiliation{Ioffe Institute, 194021 St.\,Petersburg, Russia}

\author{Tony F. Heinz}
\affiliation{Department of Applied Physics, Stanford University, Stanford, California 94305, USA and \\ 
SLAC National Accelerator Laboratory, 2575 Sand Hill Road, Menlo Park, California 94025, USA}

\author{Xavier Marie, Thierry Amand and Bernhard Urbaszek}
\affiliation{Universit\'e de
Toulouse, INSA-CNRS-UPS, LPCNO, 135 Av. Rangueil, 31077 Toulouse,
France}


\begin{abstract}
Atomically thin materials such as graphene and monolayer transition metal dichalcogenides (TMDs) exhibit remarkable physical properties resulting from their reduced dimensionality and crystal symmetry. The family of semiconducting transition metal dichalcogenides is an especially promising platform for fundamental studies of two-dimensional (2D) systems, with potential applications in optoelectronics and valleytronics due to their direct band gap in the monolayer limit and highly efficient light-matter coupling. A crystal lattice with broken inversion symmetry combined with strong spin-orbit interactions leads to a unique combination of the spin and valley degrees of freedom. 
In addition, the 2D character of the monolayers and weak dielectric screening from the environment yield a significant enhancement of the Coulomb interaction. The resulting formation of bound electron-hole pairs, or excitons, dominates the optical and spin properties of the material. Here we review recent progress in our understanding of the excitonic properties in monolayer TMDs and lay out future challenges. We focus on the consequences of the strong direct and exchange Coulomb interaction, discuss exciton-light interaction and effects of other carriers and excitons on electron-hole pairs in TMDs. Finally, the impact on valley polarization is described and the tuning of the energies and polarization observed in applied electric and magnetic fields is summarized.

\end{abstract}

\maketitle
\tableofcontents


\section{Introduction}
\label{sec:Intro}

\begin{figure*}[ht!]
\includegraphics[width=0.75\textwidth]{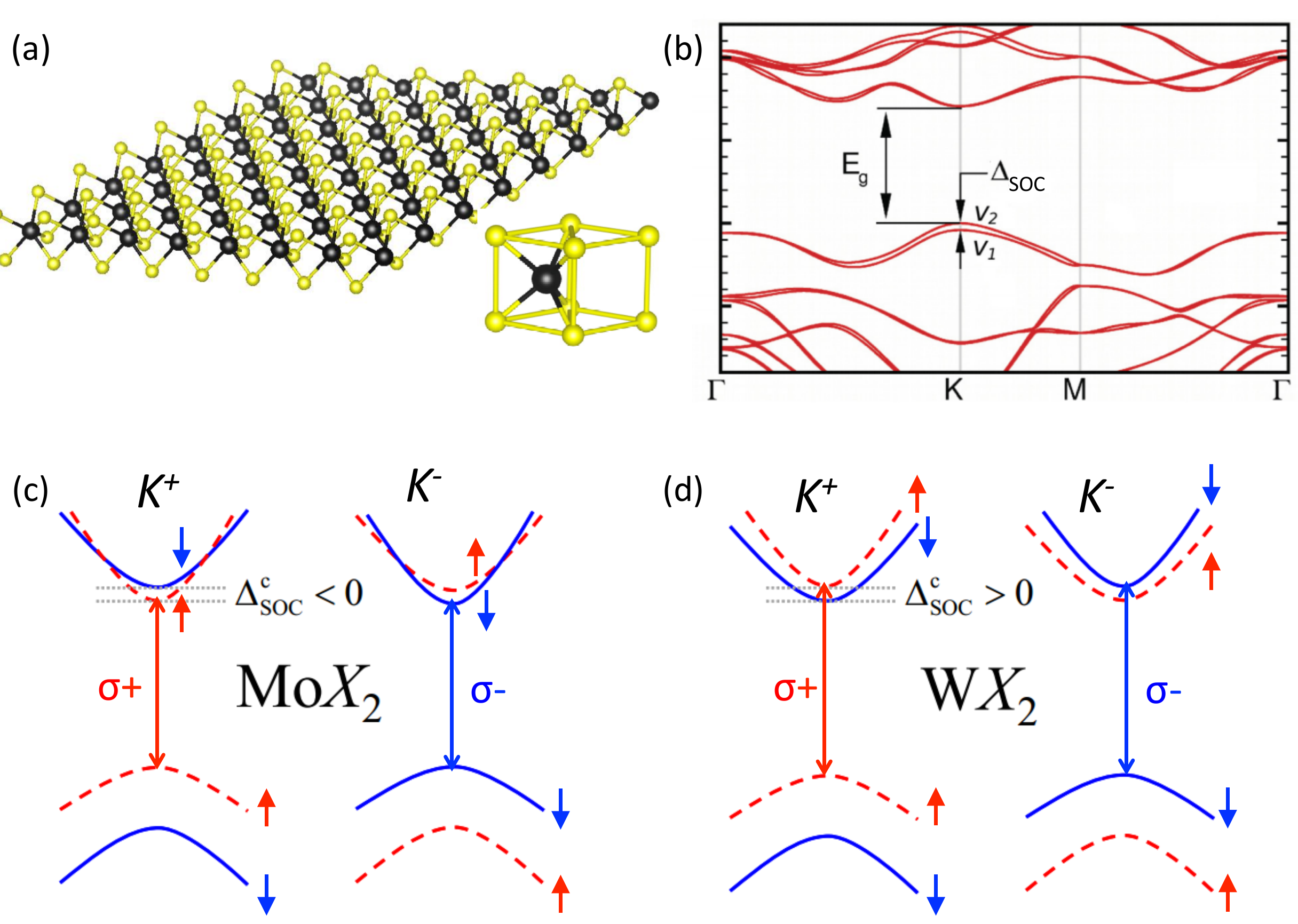}
\caption{(a) Monolayer transition metal dichalcogenide crystal structure. The transition metal atoms appear in black, the chalcogen atoms in yellow. (b) Typical band structure for MX$_2$ monolayers calculated using Density Functional Theory (DFT) and showing the quasiparticle band gap $E_g$ at the $K$ points and the spin-orbit splitting in the valence band \cite{Ramasubramaniam:2012a}. (c) Schematic in a single-particle picture showing that the order of the conduction bands is opposite in  MoX$_2$ and WX$_2$ monolayers \cite{Kormanyos:2015a}. The contribution from Coulomb exchange effects that has to be added to calculate the separation between optically active (bright - spin allowed) and optically inactive (dark - spin forbidden) excitons is not shown \cite{Echeverry:2016a}
}
\label{fig:fig1} 
\end{figure*}

Atomically thin transition metal dichalcogenides (TMDs) have unique physical properties which could be of value for a broad range of applications \cite{Wang:2012c,Geim:2013a,Butler:2013a,xia:2014,Xu:2014a,Yu:2015,Castellanos:2016a,Mak:2016a}. The investigation of bulk and thin layers of TMDs can be traced back decades ~\cite{Frindt:1966,wilson:1969,Bromley:1972a}, but starting with the emergence of graphene \cite{Novoselov:2004a,Novoselov:2005a}, many additional techniques for producing, characterizing, and manipulating atomically thin flakes were developed. This led to rapid  progress in the study of monolayers of other van der Waals systems like the TMDs. Monolayer (ML) MoS$_2$ is a typical member of the group VI TMD family of the form MX$_2$ and was isolated in early studies, for example in \textcite{Frindt:1966,Joensen:1986a}; here M is the transition metal (Mo, W) and X the chalcogen (S, Se, Te), see Fig.~\ref{fig:fig1}a. However, only around 2010, were the TMDs confirmed to be direct band gap semiconductors in monolayer form, with up to 20\% absorption per monolayer at the exciton resonance depending on the spectral region  \cite{Mak:2010a,Splendiani:2010a}. These discoveries launched intense research activity exploring the electronic properties and physics of single- and few-layer TMDs.\\
\indent The transition metal chalcogenides are a group of about 60 materials, most of which are layered structures in their bulk form with weak interlayer van-der-Waals interactions \cite{wilson:1969}. By using micro-mechanical cleavage (commonly referred to as exfoliation or ``scotch-tape technique"), one can obtain few-layer and monolayer crystals, typically a few to tens of micrometers in lateral dimension \cite{Gomez:2014a}. There are currently vigorous efforts to grow large-area TMD monolayes by chemical vapor deposition (CVD) \cite{Zhan:2012a} and by van der Waals epitaxy in ultrahigh vacuum \cite{Zhang:2014a,Xenogiannopoulou:2015a}, but many of the intriguing properties reviewed here were identified in high-quality monolayers prepared from naturally occurring or synthesized bulk crystals by exfoliation.\\
\indent In this review we mainly concentrate on group VI semiconducting dichalcogenides with M=Mo, W and X=S, Se, Te which share fascinating excitonic properties and provide unique opportunities to optically manipulate spin and valley states. These monolayers are stable enough under ambient conditions to perform optical and electrical characterization.  With respect to the electronic structure, they are indirect band gap semiconductors in their bulk form \cite{Bromley:1972a}. When thinned down to the limit of a single monolayer, the band gap becomes direct. The corresponding band extrema are located at the finite momentum $K^+$ and $K^-$ points of the hexagonal Brillouin zone and  give rise to interband transitions in the visible to near-infrared spectral range.
In the literature, the energy states close to the $K^+$/$K^-$ points located at the edges of the first Brillouin zone are typically referred to as $K^+$ and $K^-$ valleys, whereas the term valley is generally used to designate band extrema in momentum space. The presence of a direct gap is particularly interesting for potential device applications because of the associated possibility for efficient light emission.
Promising device prototypes have already been demonstrated with diverse functionality, including phototransitors based on monolayer MoS$_2$ \cite{Lopez:2013a}, sensors~\cite{Perkins:2013a}, logic circuits~\cite{Radisavljevic:2011b,Wang:2012d}, and light producing  and harvesting devices~\cite{Ross:2014a,Oriol:2014a,Cheng:2014a,Pospischil:2014a} among others.  In addition to being direct, the optical transitions at the gap are also valley selective, as $\sigma^+$ and $\sigma^-$ circularly polarized light can induce optical transitions only at the  $K^+$ and $K^-$ valleys, respectively \cite{Cao:2012a,Xiao:2012a}. 
This is in strong contrast to systems such as GaAs and many other III-V and II-VI semiconductors, where the bandgap is located at the center of the Brillouin zone ($\Gamma$-point).
In comparison to graphene, an additional interesting feature of these materials is the presence of strong spin-orbit interactions, which introduce spin splitting of several hundred meV in the valence band and of a few to tens of meV in the conduction bands \cite{Xiao:2012a,Kosmider:2013a,MolinaSanchez:2013}, where the spin states in the inequivalent valleys $K^+$ and $K^-$ are linked by time reversal symmetry.\\
\indent Since their emergence in 2010, the properties of these direct-gap monolayer materials with valley selective optical selections rules have been investigated in detail using both linear and nonlinear optical spectroscopic techniques. In a semiconductor, following absorption of a photon with suitable energy, an electron is promoted to the conduction band, leaving behind a hole in the valence band. In TMD MLs the electrons and holes are tightly bound together as excitons by the attractive Coulomb interaction, with typical binding energies on the order of 0.5\,eV  \cite{Ramasubramaniam:2012a,Cheiwchanchamnangij:2012a,Qiu:2013a,He:2014a,Chernikov:2014a,Wang:2015b}. As a result, the fundamental optical properties at both cryogenic and room temperatures are determined by strong exciton resonances. At the corresponding transition energies, the light-matter interaction is strongly enhanced in comparison to the transitions in the continuum of unbound electrons and holes. While the exciton radii are small, their properties remain to a large extent within the Wannier-Mott regime and preserve analogies to the electronic structure of the hydrogen atom. For these materials with almost ideal 2D confinement and reduced dielectric screening from the environment, the Coulomb attraction between the hole and the electrons is one to two orders of magnitude stronger than in more traditional quasi-2D systems such as GaAs or GaN quantum wells used in today's optoelectronic devices \cite{Chichibu:1996a}. Nevertheless, despite important differences, the optical properties of ML TMDs show similarities to the  exciton physics studied in detail in GaAs or ZnSe quantum wells \cite{Vinattiery,Maialle:1993a,Pelekanos:1992a,Bradford:2001a}, for example, rendering these systems a useful benchmark for comparing certain optical properties. Moreover, the Coulomb interaction in TMD MLs also determines the valley polarization dynamics of excitons and contributes to the splitting between optically \textit{bright} and \textit{dark} exciton states, in addition to spin-orbit coupling. Overall, the physics of these robust excitons are both of fundamental interest and of crucial importance for engineering and exploiting the properties of these materials in potential applications. These factors motivate this short review, which aims to present the current state of the art, as well as open questions that need to be addressed. \\
\indent The basics of the band structure and the optical spectroscopy techniques used to reveal the exciton physics in ML TMD materials are introduced in the remainder of Sec.~\ref{sec:Intro}. Neutral exciton binding energies and their impact on light-matter coupling effects are discussed in Sec.~\ref{sec:Coulomb}. Exciton physics at higher densities and in the presence of free carriers are described in Sec.~\ref{sec:complex}. Finally, the impact of the Coulomb interaction and external fields on valley physics is outlined in Sec.~\ref{sec:valley}, and open questions and challenges are addressed throughout the text to stimulate further work on the excitonic properties of atomically thin materials.
\subsection{Basic band structure and optical selection rules}
\label{sec:rules}

In addition to the strong Coulomb interaction in ML TMDs,  the crystal symmetry and orbital character of the bands are responsible for the underlying spin-valley properties and optical selection rules.  
Bulk TMDs in the semiconducting 2H phase consist of X-M-X building blocks with weak van-der-Waals bonding between the layers and are characterized by the $D_{6h}$ point symmetry group for stoichiometric compounds~\cite{wilson:1969,Ribeiro:2014a}. In bulk TMDs, the indirect band gap corresponds to the transition between the valence band maximum (VBM) at the center of the hexagonal Brillouin zone ($\Gamma$ point) and the conduction band minimum (CBM) situated nearly half way along the $\Gamma- K$ direction \cite{Zhao:2013c,Yun:2012a}. The electronic states at the $\Gamma$ point contain contributions from
the $p_z$ orbitals of the chalcogen atom and the $d_{z^2}$ orbitals of
the transition metal. In contrast, the $K^{\pm}$ point conduction
and valence band states at the corners of the hexagonal Brillouin zone
are very strongly localized in the metal atom plane, as they are
composed of transition metal atom $d_{x^2−y^2} \pm id_{xy}$ states (VB)
and $d_{z^2}$ states (CB) slightly mixed with the chalcogen $p_x \mp
i p_y$ orbitals \cite{Li:2007a, Zhu:2011a, Kormanyos:2015a}.
The spatial overlap between adjacent MX$_2$ layers of the orbitals corresponding to the $\Gamma$ point (VB) and the midpoint along $\Gamma-K$ (CB) is considerable. As a result, in progressing from bulk crystals to few layer samples and eventually to monolayers,  the indirect gap energy corresponding to the separation between $\Gamma$ and the midpoint along $\Gamma-K$
increases whereas the $K^{\pm}$ point CB and VB energies are nearly
unaffected. 
In the ML limit, the semiconductor undergoes a crossover from an
indirect to a direct gap, the later situated at the $K^{\pm}$ points (see
Fig.~\ref{fig:fig1}b)\footnote{We note that here and in the following original figures are reproduced without changes in the same format as presented in their respective source publications.}, and resulting in much stronger light emission for MLs as
compared to bulk and bilayers ~\cite{Mak:2010a,Splendiani:2010a}. \\
\indent As compared with bulk samples, the TMD MLs are described by  the lower symmetry $D_{3h}$ point group.  The symmetry elements include a horizontal $\sigma_h$ reflection plane containing the metal atoms, a threefold $C_3$ rotation axis intersecting the horizontal plane in the center of the hexagon, as well as a $S_3$ mirror-rotation axis, three twofold $C_2$ rotation axes lying in the ML plane, and mirror reflection planes $\sigma_v$ containing the $C_2$ axes~\cite{Koster63}. The symmetry of the states at $K^\pm$ is still lower and characterized by the $C_{3h}$ point group where only $C_3$, $S_3$ axes and $\sigma_h$ elements are present.\\
\indent The spin-orbit interaction in TMDs is much stronger than in graphene, the most prominent 2D material. The origin of this distinction lies simply in the relatively heavy elements in the TMDs and the involvement of the transition metal $d$ orbitals. In monolayer TMDs, the spin splitting at the $K$ point in the valence band is around 200~meV (Mo-based) and 400~meV (W-based)  \cite{Cheiwchanchamnangij:2012a,Xiao:2012a,Zhu:2011a,Zhang:2014a,Miwa:2015}. 
This coupling gives rise to the two valence sub-bands and, accordingly, to two types of excitons, A and B, which involve holes from the upper and lower energy spin states, respectively.
At the CBM, a smaller, but significant spin splitting is also expected due to partial compensation of the $p$- and $d$-states contributions~\cite{Kormanyos:2015a,Kormanyos:2014a,Liu:2013a,Kosmider:2013a}. Interestingly, depending on the metal atom (Mo or W), the conduction band spin splitting has a different sign, as indicated in Fig.~\ref{fig:fig1}c,d. 
Hence, at the \textit{K} point,
the spin degeneracy of both the conduction and valence bands is fully
lifted. This stands in marked contrast to typical GaAs or CdTe quantum-well
structures where the CBM and VBM occur at the $\Gamma$ point and both the conduction and valence band states remain spin degenerate.
The CB spin splitting results in an energy separation between the spin-allowed and optically active (bright) transitions and the spin-forbidden and optically inactive transitions (dark). The exact amplitude of the splitting for exciton states will also depend on the contribution from the electron-hole Coulomb exchange energy \cite{Qiu:2015a,Dery:2015a,Echeverry:2016a}. The lowest energy transition in MoX$_2$ is expected to be the bright exciton \cite{Kosmider:2013a,Liu:2013a}, which is consistent with temperature dependent PL measurements \cite{Wang:2015e,Zhang:2015d}, although recent studies discuss the possibility of the ground state in ML MoX$_2$ being dark \cite{Molas:2017a,Baranowski:2017a}. In contrast for the  WX$_2$ materials, dark excitons are predicted to be lower energies, in agreement with temperature dependent studies \cite{Zhang:2015d, Wang:2015e,Withers:2016,Arora:2015b}, measurements in transverse magnetic fields \cite{Zhang:2017a,Molas:2017a} and experiments probing excitons with out-of-plane dipole moments \cite{Zhou:2017a,Wang:2017b}.\\
\indent The chiral optical selection rules for interband transitions in the $K^{\pm}$ valleys can be deduced from symmetry arguments: The orbital Bloch functions of the VB states at $K^\pm$ points are invariants, while the CB states transform like the states with angular momentum components of $\pm 1$, i.e., according to the $E_1'/E_2'$ irreducible representations of the $C_{3h}$ point group. Therefore, the optical selection rules for the interband transitions at $K^\pm$ valleys are chiral: the $\sigma^+$ ($\sigma^-$) circularly polarized light can only couple to the transition at $K^+$ ($K^-$) \cite{Yao:2008, Xiao:2012a,Cao:2012a,Mak:2012a,Zeng:2012a,Sallen:2012a}. This permits the optical generation and detection of the spin-valley polarization, rendering the TMD monolayers an ideal platform to study the electron valley degree of freedom in the context of valleytronics~\cite{Schaibley:2016a,behnia:2012,Rycerz:2007a,Xiao:2007}. In that context, it is important to emphasize, that for an electron to change valley, it has either to flip its spin (see Fig.~\ref{fig:fig1}c,d) or undergo an energetically unfavorable transition, especially for the valence states. As a result, optically generated electrons and holes are \textit{both} valley and spin polarized, which is termed \textit{spin-valley locking}. Therefore, following the $\sigma^+$ excitation, the exciton emission in TMD MLs can be co-polarized with the laser if the \textit{valley} polarization lifetime is longer or of the order of the recombination time.
This behavior stands in contrast to that of III-V or II-VI quantum wells where excitation with the circularly polarized light usually results only in \emph{spin-}polarization of the charge carriers \cite{Dyakonov:2008a}. 

\begin{figure*}[ht!]
\includegraphics[width=0.85\textwidth]{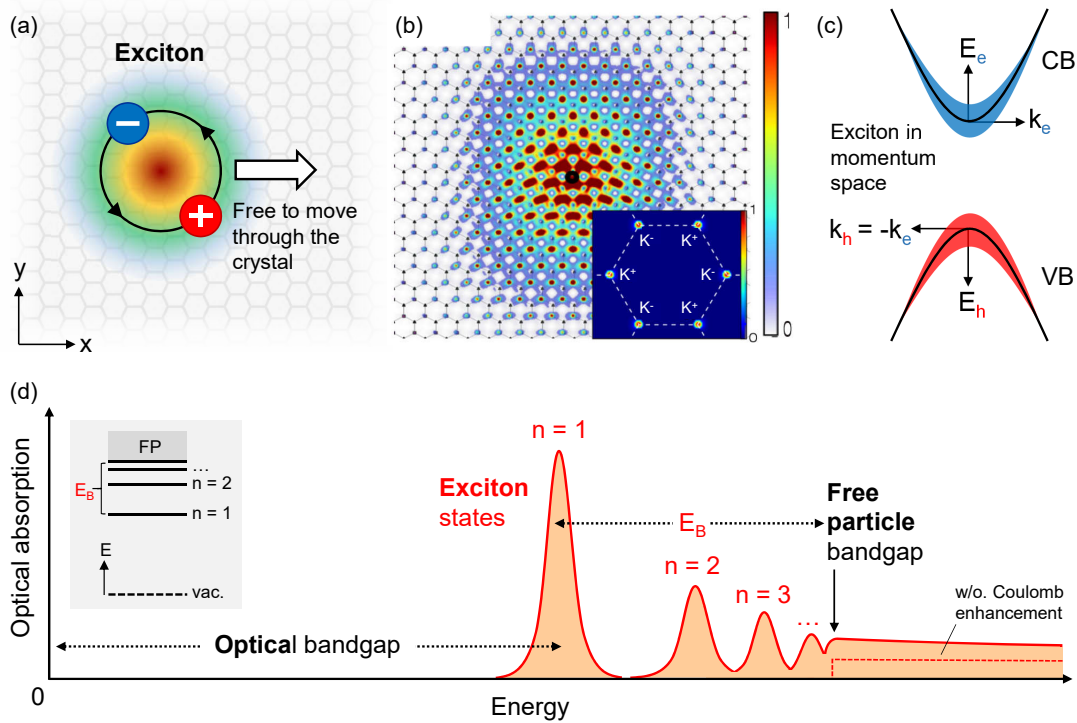}
\caption{\label{fig:fig2} 
(a) Schematic real-space representation of the electron-hole pair bound in a Wannier-Mott exciton, showing the strong spatial correlation of the two constituents. The arrow indicates the center of mass wavevector responsible for the motion of the exciton as a whole.
(b) Illustration of a typical exciton wavefunction calculated for monolayer MoS$_2$ from \cite{Qiu:2013a}. The modulus squared of the electron wavefunction is plotted in color scale (gray scale) for the hole position fixed at the origin. 
The inset shows the corresponding wavefunction in momentum space across the Brillouin zone, including both $K^+$ and $K^-$ exciton states. 
(c) Representation of the exciton in reciprocal space, with the contributions of the electron and hole quasiparticles in the conduction (CB) and valence (VB) bands, respectively, shown schematically by the width of the shaded area.
(d) Schematic illustration of the optical absorption of an ideal 2D semiconductor including the series of bright exciton transitions below the renormalized quasiparticle band gap. 
In addition, the Coulomb interaction leads to the enhancement of the continuum absorption in the energy range exceeding $E_B$, the exciton binding energy. 
The inset shows the atom-like energy level scheme of the exciton states, designated by their principal quantum number $n$, with the binding energy of the exciton ground state ($n=1$) denoted by $E_B$ below the free particle bandgap (FP)
}
\end{figure*}

\subsection{Brief survey of monolayer characterization and optical spectroscopy techniques}
\label{sec:spectro}
 
Before describing the exciton physics in detail, we summarize some relevant practical information about ML TMD samples and their typical environment (substrates) and describe the basic techniques used to investigate the optical properties. Monolayer TMDs can be obtained by the mechanical exfoliation \cite{Frindt:1966,Novoselov:2005a}, chemical exfoliation \cite{Joensen:1986a,coleman:2011a,Smith:2011a}, or CVD \cite{Liu:2012a,najmaei:2013,van:2013} and van-der-Waals epitaxy growth \cite{Xenogiannopoulou:2015a,Zhang:2014a,Liu:2015a}. Mechanical exfoliation is a convenient method to produce high-quality monolayer flakes from bulk crystals. Controlled growth of large-area monolayer material on different substrates using CVD or van-der-Waals epitaxy is a very active area of research and samples with high crystal quality have been already obtained.  \\
\indent Following isolation of a ML by micromechanical cleavage, the flakes can be deposited onto several kinds of substrates, SiO$_2$/Si, fused silica, sapphire, etc. SiO$_2$/Si substrates are widely used as (i) SiO$_2$ can help to optimize the contrast for monolayers in optical microscopy during mechanical exfoliation \cite{Lien:2015a}, and (ii) they are compatible with microelectronics standards \cite{Radisavljevic:2011a}. Encapsulation of ML flakes in hexagonal boron nitride, a layered material with a band gap in the deep UV \cite{Taniguchi:2007a}, has been shown to enhance the sharpness of the optical transitions in ML TMDs, particularly at low temperatures \cite{Chow:2017a,Jin:2016a,Manca:2017a,Cadiz:2017a,Wang:2017a,Zhou:2017a,Ajayi:2017a,Tran:2017a}. This improvement is attributed to a reduction in  detrimental surface and environmental effects on the samples.
In addition to simple optical contrast (differential reflectivity) measurements, Raman spectroscopy is often used to determine the number of layers of TMDs flakes \cite{Korn:2011a,tonndorf:2013a}. The energy spacing between two high-frequency phonon modes A$_{1g}$ and E$^1_{2g}$ can be used to identify thickness of exfoliated molybdenum dichalcogenides MX$_2$ when it is thinner than 5 layers \cite{Zhang:2015a}.  As only the monolayer is a direct-gap semiconductor (with the possible exception of MoTe$_2$ bilayers), analyzing the intensity and emission energy of photoluminescence (PL) signals allows identifying monolayer flakes. However, as the PL emission tends to favor low-energy states, including possible defect and impurity sites, care must be taken in applying this approach, especially at low temperatures. 
As an alternative, optical reflection and transmission spectroscopy can be used to identify the number of layers by quantitatively measuring the strength of the optical response. \cite{Mak:2010a,Zhao:2013b}


\section{Coulomb bound electron-hole pairs}
\label{sec:Coulomb}

In this section we summarize the main properties of the exciton states in TMD monolayers and discuss their importance for the optical response in terms of their energies (exciton resonances) and oscillator strengths (optically bright versus dark states). 
We start with a brief introduction of the electron and hole quasi-particle states forming the excitons at the fundamental band gap.
Then, we discuss the consequences of the Coulomb interaction, including direct and exchange contributions, followed by an overview of exciton binding energies and light-matter coupling in monolayer TMDs.\\
\indent The promotion of an electron from the filled valence band to the empty conduction band leaves an empty electron state in the valence band. The description of such a many-body system can be reduced to the two-particle problem of the negatively charged conduction electron interacting with a positively charged valence hole. 
The hole Bloch function $|h\rangle=|s_h,\tau_h, \bm k_h\rangle $ is derived from the Bloch function of the empty electron state $|v\rangle=|s_{v},\tau_{v},\bm k_{v}\rangle$ in the valence band by applying the time-reversal operator $|h\rangle = \hat{\mathcal K}|v\rangle$~\cite{birpikus_eng}. Here, $s_\nu$ $(\nu=c,v)$ represent the spin index, $\tau_\nu= \pm 1$ is the valley index, and $k_\nu$ is the wave vector for a conduction (c) or valence (v) state.
As the time reversal operator changes the orbital part of the wavefunction to its complex conjugate and also flips the spin, the hole wavevector is opposite that of the empty electron state, i.e., $\bm k_h=-\bm k_{v}$, the hole valley (and spin) quantum numbers are opposite to those of the empty electron state as well: $s_h=-s_{v}$, $\tau_h=-\tau_{v}$. This transformation is natural to describe the formation of the electron-hole pair from the photon with a given polarization.
In case of TMD monolayers, a $\sigma^+$ photon with a wavevector projection $\bm q_\parallel$ to the plane of the layer creates an electron with a wavevector $\bm k_e$ in the $s_e=+1/2$ state in $\tau_e=+1$ ($K^+$) valley, leaving a state with wavevector $\bm k_{v}=\bm k_e - \bm q_\parallel$ in the valence band unoccupied.
As a result, the corresponding hole wavevector is $\bm k_h=-\bm k_v={\bm q_\parallel} -\bm k_e$, with the center of mass wavevector of the electron-hole pair equal to $\bm K_\text{exc}= \bm k_e + \bm k_h={\bm q_\parallel}$, as expected for the quasiparticle created by a photon. Accordingly, the hole valley index, $\tau_h=-1$, and spin, $s_h=-1/2$, are formally opposite to those of the conduction-band electron. 
In a similar manner, the absorption of $\sigma^-$ photon results in the formation of the electron-hole pair with $\tau_e=-\tau_h=-1$, $s_e=-s_h=-1/2$~\cite{Glazov:2014a,PSSB:PSSB201552211}.

\subsection{Neutral excitons: direct and exchange Coulomb interaction}
\label{sec:diffTerm} 

To discuss the consequences of the Coulomb electron-hole interaction we separate the \textit{direct} and \textit{exchange} contributions, both further including \textit{long-range} and \textit{short-range} interactions, with certain analogies to traditional quasi-2D quantum well excitons \cite{Dyakonov:2008a}.
The \textit{long-range} part represents the Coulomb interaction acting at inter-particle distances in real space larger than the inter-atomic bond lengths (i.e., for small wavevectors in reciprocal space compared to the size of the Brillouin zone). 
In contrast, the \textit{short-range} contribution originates from the overlap of the electron and hole wavefunctions at the scales on the order of the lattice constant ($a_0\simeq 0.33$~nm in ML WSe$_2$), typically within one or several unit cells (i.e., large wavevectors in reciprocal space).\\
\indent The \textit{direct} Coulomb interaction describes the interaction of positive and negative charge distributions related to the electron and the hole.
The \textit{long-range} part of the \textit{direct} interaction is determined mainly by the envelope function of the electron-hole pair being only weakly sensitive to the particular form of the Bloch functions, i.e., valley and spin states; it rather depends on the dimensionality and dielectric properties of the system. 
It has an electrostatic origin and provides the dominant contribution to the exciton binding energy, $E_B$, see Sec.~\ref{sec:Ebind}. 
The \textit{short-range} part of the direct interaction stems from the Coulomb attraction of the electron and the hole within the same or neighboring unit cells. It is sensitive to the particular form of the Bloch functions and is, as a rule, considered together with the corresponding part of the exchange interaction.
In a semi-classical picture, the long-range direct interaction thus corresponds to attractive Coulomb forces between opposite charges. 
As a consequence, an electron and a hole can form a bound state, the neutral exciton, with strongly correlated relative positions of the two constituents in real space, as schematically shown in Fig.~\ref{fig:fig2}a.
The concept of correlated electron-hole motion is further illustrated in Fig.~\ref{fig:fig2}b, as reproduced from Ref.\,\cite{Qiu:2013a}, where the modulus squared of the electron wavefunction relative to the position of the hole is presented for the case of the exciton ground state in monolayer MoS$_2$. In TMD MLs, the exciton Bohr radius is on the order of one to a few nanometers and the correlation between an electron and a hole extends over several lattice periods.
Thus, strictly speaking, the exciton could be formally understood to be of an intermediate nature between the so-called \textit{Wannier-Mott} or large-radius type similar to prototypical semiconductors such as GaAs and Cu$_2$O and \emph{Frenkel} exciton, which corresponds to the charge transfer between nearest lattice sites. However,  to describe the majority of the experimental observations, the Wannier-Mott description in the effective mass approximation appears to be largely appropriate even for quantitative predictions.
\\
\indent In the $\bm k$-space, the exciton wavefunction $\Psi^{X}$ can be presented as~\cite{birpikus_eng,PSSB:PSSB201552211}
\begin{equation}
\label{psi}
\Psi{^{X}} = \sum_{e, h}C{^{X}}(\bm k_e,\bm k_h)|e; h\rangle,
\end{equation}
where the correlation of the electron and hole in the exciton is described by a coherent, i.e., phase-locked, superposition of electron and hole states ($|e\rangle=|s_e,\tau_e, \bm k_e\rangle$ and $|h\rangle=|s_h,\tau_h, \bm k_h\rangle$) around the respective extrema of the bands. Relative contributions of these states to the exciton are described by the expansion coefficients $C^{X}$, which are usually determined from the effective two-particle Schr\"{o}dinger or Bethe-Salpeter equation.
Their values are schematically represented by the size of the filled area in Fig.~\ref{fig:fig2}c, with the results of an explicit calculation shown in the inset of Fig.~\ref{fig:fig2}b for electrons in monolayer MoS$_2$. As a consequence of the large binding energy of excitons and their small Bohr radius in real-space ($a_B\simeq 1$~nm), the spread of the exciton in $\bm k$-space is significant. Therefore states further away from the $K$-point band extrema (see inset in Fig. 2b) are included in the exciton wavefunction \cite{Wang:2015c,Qiu:2013a}. \\ 
\indent As previously noted, the correlation represented in Eq.\,\eqref{psi} is strictly related to the relative motion of the carriers.
In contrast, the exciton center-of-mass can propagate freely in the plane of the material, in accordance with the Bloch theorem.
The resulting exciton states $X=\{\bm K_\text{exc}, s_e, \tau_e, s_h, \tau_h, (n, m)\}$ are labeled by the center-of-mass wavevector $\bm K_\text{exc}$, electron and hole spin and valley indices, $s_e$, $\tau_e$, $s_h$, $\tau_h$ and the relative motion labels $(n,m)$.
The relative motion states can be labelled by the principal and magnetic quantum number as $(n,m)$, with $n=1,2,3 ...$ a natural number, $m \in \mathbb{Z}$ and $|m|<n$. To choose a notation similar to the hydrogen atom for $s,p,d$ states, we use here $(n,0)=ns$ where $n \in  \mathbb{N}$ and $(n,\pm1)=(np,\pm1)$ for $n>1$, $(n,\pm2)=(nd,\pm2)$ for $n>2$ etc; the precise symmetry of excitonic states is discussed below in Sec.~\ref{sec:light}.

In particular, the principal quantum number $n$ is the primary determinant of the respective binding energy, with the resulting series of the ground state ($n=1$) and excited states ($n>1$) of Wannier-Mott excitons roughly resembling the physics of the hydrogen atom, as represented by the energy level scheme in Fig.~\ref{fig:fig2}d. The selection rules for optical transitions are determined by the symmetry of the excitonic wavefunctions, particularly, by the set of the spin and valley indices $s_{e, h}$ and $\tau_{e, h}$ and the magnetic quantum number $m$.
These quantities are of particular importance for the subdivision of the excitons into so-called \textit{bright} states, or optically active, and \textit{dark} states, i.e., forbidden in single-photon absorption process, as further discussed in the following sections. \\
\indent In addition to the formation of excitons, a closely related consequence of the Coulomb interaction is the so-called \textit{self-energy} contribution to the absolute energies of electron and hole quasiparticles.
In a simplified picture, the self energy is related to the repulsive interaction between identical charges and leads to an overall increase of the quasiparticle band gap of a semiconductor, i.e., the energy necessary to create an unbound electron-hole pair in the continuum, referred to as 'free-particle (or quasiparticle) band gap'.  
In many semiconductors, including TMD monolayers, the self-energy contribution and the exciton binding energy are found to be almost equal, but of opposite sign.
Thus, the two contributions tend to cancel one another out with respect to the absolute energies.

Nevertheless, these interactions are of central importance as they determine the nature of the electronic excitations and the resulting properties of the material.
To demonstrate the later, a schematic illustration of the optical absorption in an ideal 2D semiconductor is presented in Fig.~\ref{fig:fig2}d.
The changes associated with the presence of strong Coulomb interactions result in the formation of the exciton resonances below the renormalized free-particle band gap.
Importantly, the so-called \textit{optical} band gap is then defined with respect to the lowest energy feature in absorption, i.e., the ground state of the exciton ($n=1$).
The optical gap thus differs from \textit{free-particle} band gap, which corresponds, as previously introduced, to the onset of the continuum of unbound electrons and holes. The free particle bandgap is thus formally equivalent to the $n=\infty$ limit of the bound exciton state.
Consequently, samples with different exciton binding energies and free particle bandgaps can have \emph{optical} bandgaps at very similar energies. This can be illustrated, e.g., in comparative studies of the absolute energies of exciton resonances for monolayer samples placed in different dielectric environments, effectively tuning both the exciton binding energy and the free particle bandgap~\cite{Stier:2016b, Raja:2017a, Cadiz:2017a}.
As a final point, the Coulomb interaction leads to a significant enhancement of the continuum absorption, which is predicted to extend many times of $E_B$ into the band\,\cite{Shinada:1966a,Haug2009}.\\
\indent In comparison to the \textit{direct} coupling part of the Coulomb interaction, the \textit{exchange} contribution denotes the Coulomb interaction combined with the Pauli exclusion principle. The latter is a well-known consequence of the fact that both types of quasiparticles (electrons and holes) result from a sea of indistinguishable charged fermions occupying filled bands: The wavefunction of the many-electron system with one carrier promoted to the conduction band should be antisymmetrized with respect to permutations of the particles. Hence, just as for exchange interaction in atoms, the energy of exciton depends on the mutual orientation of electron and hole spins and, as a particular feature of the TMD MLs, on the quasi-particle valley states.
In analogy to the direct coupling, the Coulomb exchange interaction can be also separated into the long-range and the short-range parts. 
In particular, the \textit{long-range} exchange interaction is of electrodynamic nature, in close analogy to the exchange interaction between an electron and a positron~\cite{positronium}. It can be thus interpreted as a result of interaction of an exciton with the induced electromagnetic field in the process of virtual electron-hole recombination~\cite{birpikus_eng,denisovmakarov,goupalov98}: The bright exciton can be considered as a microscopic dipole which produces an electric field, the back-action of this field on the exciton is the long-range electron-hole exchange interaction. On a formal level, it corresponds to the decomposition of the Coulomb interaction up to the dipole term and calculation of its matrix element on the antisymmetrized Bloch functions~\cite{andreani:book}. 
In TMD monolayers, the \textit{long-range} exchange part, being much larger than for III-V or II-VI quantum wells, facilitates transitions between individual exciton states excited by the light of different helicity, thus mainly determining the spin-valley relaxation of the excitons, see Sec.~\ref{sec:valley}.
At \textit{short-range}, Pauli exclusion causes the exchange interaction to depend strongly on the spin and valley states of the particles. 
It thus contributes to the total energies of the many-particle complexes, depending on the spin and valley states of the individual constituents and impacts the separation between optically dark and bright excitons \cite{Qiu:2015a,Echeverry:2016a}.
Among typical examples are the so-called triplet and singlet exciton states (i.e., the exciton fine structure) corresponding to parallel and anti-parallel alignment of the electron and hole spins, respectively. 
Lacking a classical analog, the exchange interaction is a more subtle contribution compared to the direct Coulomb interaction. A rough estimate of the exchange to direct interaction ratio in exciton is provided by the ratio of the binding energy, $E_B$, and the band gap, $E_g$: $\sim E_B/E_g$~\cite{birpikus_eng}.
As it is summarized in Table\,\ref{tab:table1}, the overall ratio of the direct and exchange contributions in TMDs is on the order of $10:1$, depending, in particular, on the exciton wavevector for the long-range interaction~\cite{Glazov:2014a}.
Nevertheless, as it is discussed in the following sections, the consequences of exchange interaction are of central importance in understanding many-particle electronic excitations in TMD monolayers.

\begin{table}
\caption{\label{tab:table1} Impact of different types of electron-hole interaction on optical and polarization properties of excitons in TMD MLs.}
\begin{ruledtabular}
\begin{tabular}{cc}  
  Coulomb term & Impact  \\
 \hline
\textbf{Direct}  & \textbf{Exciton binding energy}\\
& neutral excitons $\sim 500$~meV\\
& charged excitons, biexcitons $\sim 50$~meV \\
& \textbf{Quasi-particle bandgap}  \\
& self-energy $\sim 500$~meV\\
\hline
\textbf{Exchange} & \textbf{Exciton fine structure}\\
long-range  & neutral exciton spin/valley depolarization  \\
 & $\sim 1\ldots 10$ meV \\
short-range & splitting of dark and bright excitons \\
& $\sim 10$`s of~meV \\
\end{tabular}
\end{ruledtabular}
\end{table} 

One of the distinct properties of TMD monolayers is the unusually strong long-range Coulomb interaction and its unconventional interparticle distance dependence, leading to large exciton binding energies and band-gap renormalization effects.
First, the decrease of dimensionality results in smaller effective electron and hole separations, particularly, perpendicular to the ML plane, where the wavefunctions of the electron and hole occupy only several angstroms as compared to tens of nanometers in bulk semiconductors. In the simple hydrogenic model, this effect yields to a well-known four-fold increase in exciton binding energy in 2D compared to 3D ~\cite{Ivchenko:2005a}.  Second, the effective masses in the $K^\pm$ valleys of the electron, $m_e$, and hole, $m_h$, in TMD MLs are relatively large, on the order of $\sim 0.5\,m_0$, with $m_0$ denoting the free electron mass~\cite{Liu:2013a,Kormanyos:2015a}. Hence, the reduced mass $\mu=m_em_h/(m_e+m_h)\approx 0.25~m_0$ is also larger compared to prominent semiconductor counterparts such as GaAs ($\mu \approx 0.06~m_0$). Finally, in TMD MLs, the material is generally surrounded by air/vacuum (or dielectrics with relatively small permitivity). This reduces dielectric screening of the Coulomb interaction, since the electric field produced by the electron-hole pair is present largely outside of the ML itself. These features of the screening also result in a substantial deviation of the electron-hole interaction from the conventional $1/r$ distance dependence, as discussed in detail in Sec.~\ref{sec:env}. Nevertheless, one can still estimate the impact of the dimensionality, the effective mass, and the reduced screening on the exciton binding energy $E_B$ within the framework of the 2D hydrogen-like model:  $E_B \propto 4Ry~\mu /(m_0\varepsilon_{\rm eff}^2)$, where $Ry$ is the Rydberg constant of 13.6\,eV and $\varepsilon_{\rm eff}$ is a typical effective dielectric constant of the system, roughly averaged from the contributions of the ML and the surroundings, $m_0$ is the free electron mass. Clearly, an increase in $\mu$ and decrease in $\varepsilon_{\rm eff}$ result in the increase of the binding energy. As an example, this simple expression provides a binding energy on the order of 0.5~eV for realistic parameters of $\mu=0.25~m_0$ and $\varepsilon_{\rm eff} = 5$.

As a final step in introducing the Coulomb terms and their role in the physics of TMD monolayers, we can formally identify the direct and exchange terms in the effective exciton Hamiltonian in $\bm k$-space in the two-band approximation:
\begin{multline}
\label{HX}
\mathcal H_{XX'}(\bm k_e, \bm k_h;\bm k_e',\bm k_h') =\\
\left[ \mathcal H_e(\bm k_e) \delta_{\bm k_e,\bm k_e'} + \mathcal H_h(\bm k_h)   \delta_{\bm k_h,\bm k_h'} + V_{\bm k_e\bm k_h;\bm k_e',\bm k_h'} \right] \delta_{XX'} +\\ U_{\bm k_e\bm k_h;\bm k_e',\bm k_h'}(EH;E'H') \delta_{\bm K, \bm K'},
\end{multline}
where $\mathcal H_e(\bm k_e)$ ($\mathcal H_h(\bm k_h)$) are the electron (hole) single-particle Hamiltonians, $V_{\bm k_e\bm k_h;\bm k_e',\bm k_h'}$ stands for the matrix element of the direct (long-range) Coulomb interaction between the electron and the hole, and $U_{\bm k_e\bm k_h;\bm k_e',\bm k_h'}(EH;E'H')$ is the matrix of the electron-hole exchange interaction. Here $E=s_e\tau_e$, $H=s_h\tau_h$ are the electron and hole spin and valley indices, the dependence of the single-particle Hamiltonians on $E$ and $H$ is implicitly assumed. The last term comprises the short- and long-range contributions to the electron-hole exchange interaction. In real space, the second line of Eq.~\eqref{HX} corresponds to the standard exciton Hamiltonian in the effective mass approximation with a properly screened Coulomb interaction potential with the additional short-range part in the form $V_0(EH;E'H')\delta(\bm r_e - \bm r_h)$ with the parameters $V_0(EH;E'H')$ determined by particular form of the Bloch functions~\cite{birpikus_eng}. 

\subsection{Exciton binding energy}
\label{sec:Ebind} 

\subsubsection{Exciton and continuum states in optics and transport}
\label{sec:andcont} 

\begin{figure*}[ht!]
\includegraphics[width=1\textwidth]{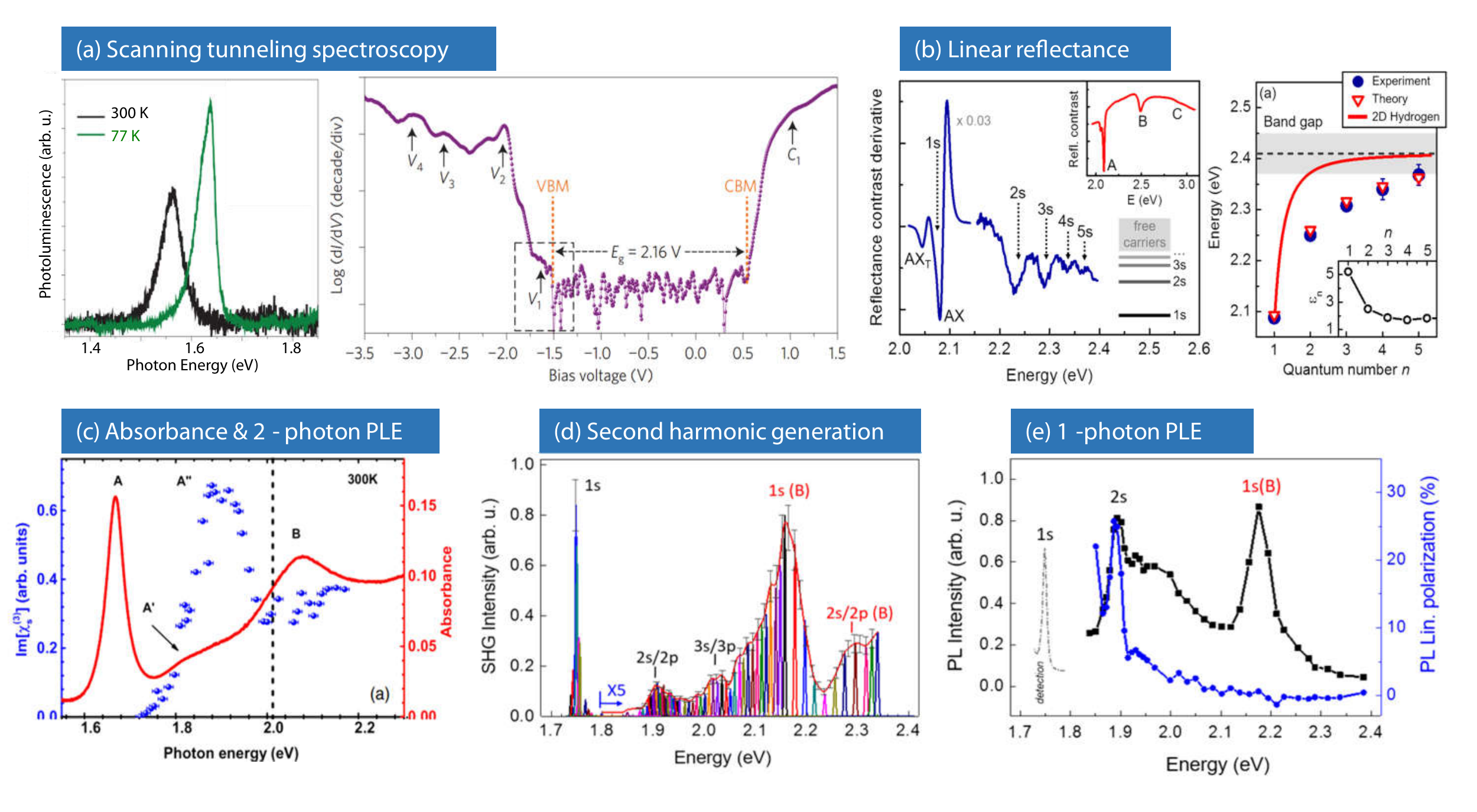}
\caption{\label{fig:fig3} Presentation of commonly used experimental techniques to determine exciton binding energies in TMD monolayers. (a) Direct measurement of the free-particle bandgap energy using scanning tunneling spectroscopy of ML MoSe$_2$ on bilayer graphene (right panel) combined with a measurement of the absolute energy of the exciton ground state from photoluminescence (left panel)~\cite{Ugeda:2014a}. (b) Exciton states of ML WS$_2$ on an SiO$_2$/Si substrate from reflectance contrast measurements~ \cite{Chernikov:2014a}. The extracted transition energies of the states and the inferred band-gap position are presented in the right panel.  (c) The linear absorption spectrum and the third-order susceptibility extracted from two-photon photoluminescence excitation spectra of ML WSe$_2$ on fused silica substrate with exciton resonances of the ground and excited states~\cite{He:2014a}. (d) Exciton states as measured by second-harmonic spectroscopy of the A and B transitions in ML WSe$_2$~ \cite{Wang:2015b}. (e) One-photon photoluminescence excitation spectra and the degree of linear polarization of the luminescence of ML WSe$_2$ with features of excited $2s$ state of the A and the ground state of the B exciton~ \cite{Wang:2015b}.}

\end{figure*}

To determine the exciton binding energy $E_B$ directly by experiment, one must identify both the absolute energy position of the exciton resonance $E_X$ and that of the free-particle bandgap $E_{g}$ to obtain $E_B=E_{g}-E_X$.
For this purpose, several distinct techniques have been successfully applied to TMD monolayers.
The transition energy $E_X^{(n=1)}$ of the exciton ground state can be readily obtained using optical methods.
Due to the strong light-matter coupling (cf. Sec.~\ref{sec:light}) the excitons appear as pronounced resonances centered at photon energies corresponding to $E_X^{(n=1)}$ in optical absorption, reflectance, photoluminescence (PL), photoluminescence excitation (PLE), and photocurrent (PC) measurements. In case of PL, room-temperature measurements are usually preferred to avoid potential contributions from defect states.
As an example, PL spectra of MoSe$_2$ monolayer from Ref.~\cite{Ugeda:2014a} are presented in the left panel of Fig.~\ref{fig:fig3}a, illustrating the strong emission from the ground-state exciton transition.\\
\indent In contrast, the precise determination of the free-particle bandgap energy is more challenging problem – and a recurring one for semiconductors with large exciton binding energies where strong exciton resonances may mask the onset of a continuum of states.
A direct approach is provided by the scanning tunneling spectroscopy (STS), which
measures tunneling currents as a function of the bias voltage through a tip positioned in close proximity to the sample.
Such measurements can probe the electronic density of states in the vicinity of the band gap, mapping energy levels of free electrons in both the valence and conduction bands.
A typical STS spectrum for a MoSe$_2$ monolayer supported by bilayer of graphene~\cite{Ugeda:2014a} is presented in the right panel of Fig.~\ref{fig:fig3}a.
As a function of tip voltage relative to the sample, a region of negligible tunneling current is observed. This arises from the band gap where no electronic states are accessible.
The lower and upper onset of the tunnel current correspond to the highest occupied electron states at the VBM and the lowest unoccupied states at the CBM, respectively.
The size of the bandgap $E_{g}$ is extracted from the difference between these onsets.
As previously discussed, the exciton binding energy is then directly obtained from the difference between $E_{g}$ measured by STS and the exciton transition energy $E_X^{(n=1)}$ identified in the optical spectroscopy (compare right and left panel in Fig.~\ref{fig:fig3}a).
The reported values, as summarized in the Table \ref{tab:table2}, range from 0.22 eV for MoS$_2$~\cite{Zhang:2014d} to 0.55 eV for MoSe$_2$~\cite{Ugeda:2014a}; further reports include \cite{Bradley:2015a, chiu:2015, Rigosi:2016a, Zhang:2015c}.
The differences can be related to (i) the overall precision in extracting the onsets of the tunneling current and (ii) to the use of different conducting substrates required for the STS, i.e., the influence of different dielectric environments and related proximity effects.
In addition, the complexities of the band structure of the TMDs, with several valley extrema being relatively close in energy (see Sec.~\ref{sec:rules}) were shown to be of particular importance for the identification of the bands contributing to the initial rise in the tunneling current ~\cite{Zhang:2015c}. 

As discussed in Sec.~\ref{sec:diffTerm} (see Fig.~\ref{fig:fig2}d), the onset of the free-particle continuum in the absorption spectra is merged with the series of excited exciton states ($n=2,3,...$), precluding a direct extraction of the bandgap energy in most optical spectroscopy experiments.
However, the identification of the series of excited exciton states permits an extrapolation to the expected band gap or for the determination of the band gap through the application of suitable models.
These methods are analogous to the measurements of the Rydberg (binding) energy of the hydrogen atom from spectral lines from transitions between different electron states. For an ideal 2D system the exciton energies evolve as $E_{B}^{n} = \mu e^4/[2\hbar^2\varepsilon_{\rm eff}^2 (n-1/2)^2]$ in a hydrogenic series with $n=1,2...$ \cite{Shinada:1966a,Klingshirn:2007}. As clearly shown in reflection spectroscopy \cite{Chernikov:2014a,He:2014a}, the exciton states in ML WSe$_2$ and WS$_2$, for example, deviate from this simple dependence, see Fig.~\ref{fig:fig3}b. The main reason for the change in the spectrum is the nonlocal dielectric screening associated with the inhomogeneous dielectric environment of the TMD ML. This results in a screened Coulomb potential ~\cite{Cudazzo:2011a,1979JETPL..29..658K,Rytova:1967} with a distance dependence that deviates strongly from the usual $1/r$ form, as detailed below, and also introduced in the context of carbon nanotubes~\cite{Wang:2005a,1742-6596-129-1-012012,Deslippe:2009a,Ando:2010aa}.

The energies of the excited states of the excitons $n>1$ can be directly obtained from linear absorption or reflectance spectroscopy.
These states are usually identified by their decreasing spectral weight (oscillator strength) and relative energy separations with increasing photon energies. The oscillator strength for an ideal 2D system is given by $f_n=f_{n=1}/(2n-1)^3$ \cite{Shinada:1966a}.
As an example, consider the reflectance contrast spectrum (i.e., the difference of the reflectity of the sample and substrate divided by the substrate reflectivity) from a WS$_2$ monolayer \cite{Chernikov:2014a}, measured at cryogenic temperatures.  The spectrum, presented after taking a derivative with respect to photon energy to highlight the features in the left panel of Fig.~\ref{fig:fig3}b, reveals signatures of the excited exciton states.
The right panel summarizes the extracted peak energies and the estimated position of the band gap, as obtained directly from the extrapolation of the data and from model calculations.  The corresponding exciton binding energy is about 300\,meV.
Observations of the excited states in reflectance spectra were further reported for WSe$_2$~\cite{He:2014a,Hanbicki:2015a,Arora:2015b} and WS$_2$ \cite{Hanbicki:2015a,Hill:2015a} monolayers, both at cryogenic and room temperature, as well as for MoSe$_2$ \cite{Arora:2015c}. 
In addition, the relative energy separations between the ground and excited states of the excitons were found to decrease with thickness of multilayer samples \cite{Chernikov:2014a,Arora:2015b}, reflecting the expected decrease in the binding energy.
Similar results were obtained by the related techniques of photoluminescence-excitation spectroscopy (PLE) \cite{Hill:2015a,Wang:2015b} and photocurrent (PC) \cite{Klots:2014a} spectroscopy, which also allow identification of the ground and excited-state excitonic transitions.
In both cases, this is achieved by tuning the photon energy of the excitation light source, while the luminescence intensity of a lower-lying emission feature (in PLE) or the current from a sample fabricated into a contacted device (in PC) are recorded. PLE is a multistep process: light is first absorbed, then energy relaxation occurs to the optically-active $1s$ exciton. As relaxation via phonons plays an important role in TMD MLs \cite{Molina:2011a,Chow:2017b}, the PLE spectra contain information on both absorption and relaxation pathways. From PLE measurements, excited states of the excitons were observed in WSe$_2$ \cite{Wang:2015b}, WS$_2$ \cite{Hill:2015a}, MoSe$_2$ \cite{Wang:2015c} and MoS$_2$ \cite{Hill:2015a} monolayers.
In PC, the onset of the bandgap absorption in MoS$_2$ monolayers was reported in Ref. \textcite{Klots:2014a}.\\
\indent One of the challenges for linear absorption or reflectance spectroscopy, is the dominant response from the exciton ground state, potentially obscuring weaker signatures from the excited states.
As an alternative, excited states of the excitons for example $(n,\pm1)=(np,\pm1)$ for $n>1$ can be addressed via two-photon excitation in TMDs~\cite{Berkelbach:2015,Wang:2015b,2p:1,Ye:2014a}, while the two-photon absorption by the dipole-allowed transitions for $(n,0)=1s,2s,3s...$ is strongly suppressed. 
Indeed, in the standard centrosymmetric model $s$-shell excitons are allowed in one-photon processes (and forbidden in all processes involving even number of phonons), while $p$-shell excitons are allowed in two-photon processes and forbidden in one-photon processes~\cite{Mahan:1968ly}. Note that the specific symmetry of the TMD ML can lead to a mixing between exciton $s$ and $p$-states and activation of $p$-states in single-photon transitions as well \cite{Glazov:2017a,Gong:2017a}. The mixing is also proposed to originate from a small amount of disorder in the system \cite{Berghauser:2017a} \\
\indent Here, a commonly used technique is two-photon photoluminescence excitation spectroscopy (2P-PLE).
In this method, the (pulsed) excitation source is tuned to half the $p$ exciton transition energy and the resulting luminescence is recorded as a function of the photon excitation energy.   
Formally, this yields the spectrum of third-order nonlinear susceptibility responsible for two-photon absorption. The result of such a 2P-PLE measurement of a WSe$_2$ monolayer \cite{He:2014a} is presented in Fig.~\ref{fig:fig3}c.In contrast to one-photon absorption, the two-photon response is dominated by resonances from the excited exciton states with $p$-type symmetry, such as the 2$p$, 3$p$ ... states of the A-exciton (labelled A$^{\prime}$ and A$^{\prime\prime}$ in Fig.~\ref{fig:fig3}c).
Further reports of the exciton states in 2D TMDs from 2P-PLE include studies of WS$_2$ \cite{Ye:2014a, Zhu:2015b}, WSe$_2$ \cite{Wang:2015b} and MoSe$_2$ monolayers \cite{Wang:2015c}.
Like the analysis of the one-photon spectra, the band gap is extracted either by comparison of the ground and excited state energies with appropriate theoretical models \cite{Ye:2014a, Wang:2015b} or from the estimated onset of the continuum absorption (free-particle gap) \cite{He:2014a, Zhu:2015b}.
In addition to the PLE experiments, both the ground and excited states can be also observed directly in second-harmonic generation spectra, as illustrated in Fig.~\ref{fig:fig3}d for WSe$_2$ monolayers \cite{Wang:2015b}. 
The second-harmonic generation takes place because, due to the lack of an inversion center in TMD MLs, the  $s$-shell and $p$-shell excitons become active both in single- and two-photon processes. This allows for excitation of the given exciton state by two photons and its coherent emission.  
The microscopic analysis of the selection rules and relative contributions of excitonic states in second-harmonic emission is presented in Ref.~\cite{Glazov:2017a}, see also~\cite{PhysRevB.89.235410}.
Overall, the main challenge with optical techniques is the correct identification of observed features, made more challenging by a the possible mixture of $s$ and $p$ excitons, as well as coupling to phonon modes \cite{Jin:2016a,Chow:2017a}.
Topics of current discussion in analyzing different spectra include possible contributions from phonon-assisted absorption, higher-lying states in the band structure, defects, and interference effects.\\
\indent Further information on exciton states and their energy can be obtained from measurements of intra-exciton transitions in the mid-IR spectral range after optical injection of finite exciton densities \cite{Poellmann:2015, Cha:2016} and measurements of the exciton Bohr radii from diamagnetic shifts at high magnetic fields \cite{Stier:2016a, Stier:2016b}.
The latter approach also allows for the estimation of the binding energy of the B-exciton related to the spin-orbit split valence subband.
A summary of the exciton binding energies and the corresponding band-gap energies is presented in Table \ref{tab:table2}.
While the extracted absolute values vary, largely due to the outlined challenges of precisely determining the absolute position of the band gap, the following observations are compatible with the majority of the literature:\\
(1) Excitons are tightly bound in TMD monolayers due to the quantum confinement and low dielectric screening, with binding energies on the order of several 100's of meV. 
The corresponding ground-state Bohr radii are on the order of nanometers and the wavefunction extends over several lattice constants $a_0$ (for WSe$_2$ $a_0 \approx 0.33$~ nm), rendering the Wannier-Mott exciton model largely applicable.\\
(2) The absolute position of the free-particle bandgap renormalizes by an amount similar to the exciton binding energy in comparison to the respective $K-K$ transition in bulk.  Thus, we observe  only to a modest absolute shift of the exciton energy in optical spectra when comparing the bulk and monolayers.\\ 
(3) The Coulomb interaction deviates from the $1/r$ law due to the spatially inhomogeneous dielectric screening environment (see Sec.~\ref{sec:env}).
This changed distance dependence of the $e-h$ interaction strongly affects the energy spacing of the $n=1,2,3...$ exciton states, leading to pronounced deviations from the 2D hydrogen model.

\begin{table*}[ht]
\caption{Summary of experimentally determined exciton binding energies and free particle bandgaps in monolayer TMDs from the literature. 
	All values correspond to the A-exciton transition, unless noted otherwise. The numerical formats correspond to the presentations of the data in the respective reports.}
  \centering
	\begin{tabular}{c|c|c|c|c|c}
    \textbf{Material} & \textbf{Sample (Temp.)} & \textbf{Exp. technique} & \textbf{~Bind. energy [eV]~} & \textbf{Bandgap} [eV]& \textbf{Reference}\\
    \hline
		\hline
    WSe$_2$		& Exf. on SiO$_2$/Si (RT) 				& Refl., 2P-PLE  			& 0.37 						& 2.02 							& \onlinecite{He:2014a}\\
							& CVD on HOPG	(79\,K)		 					& STS, PL	 						& 0.5 						& 2.2$\pm$0.1 			& \onlinecite{Zhang:2015c}\\
							& Exf. on SiO$_2$/Si (4\,K)				& ~PLE, 2P-PLE, SHG~ 	& 0.6$\pm$0.2			& 2.35$\pm$0.2 			& \onlinecite{Wang:2015b}\\
							& Exf. on SiO$_2$/Si (4, 300\,K)	& Refl. 							& 0.887						& 2.63				 			& \onlinecite{Hanbicki:2015a}\\
							& CVD on HOPG (77\,K)			 				& STS, PL 		 				& $\approx$\,0.4*	& 2.08$\pm$0.1 			& \onlinecite{chiu:2015}\\
							& Exf. on diamond$_2$ (RT)			 			& Mid-IR pump-probe 		 	& 0.245   & 1.9\,***	& \addAlexey{\onlinecite{Poellmann:2015}}\\
							
		\hline
		WS$_2$		& Exf. on SiO$_2$/Si (5\,K)				& Refl. 							& 0.32$\pm$0.04		& 2.41$\pm$0.04			& \onlinecite{Chernikov:2014a}\\
							& Exf. on fused silica (10\,K)		& 2P-PLE 		 					& 0.7 						& 2.7					 			& \onlinecite{Ye:2014a}\\
							& Exf. on SiO$_2$/Si (RT)					& 2P-PLE 				 			& 0.71$\pm$0.01		& 2.73				 			& \onlinecite{Zhu:2014b}\\
							& Exf. on SiO$_2$/Si (4, 300\,K)	& Refl. 							& 0.929						& 3.01				 			& \onlinecite{Hanbicki:2015a}\\
							& Exf. on fused silica (RT)				& Refl., PLE					& 0.32$\pm$0.05		& 2.33$\pm$0.05			& \onlinecite{Hill:2015a}\\
							& Exf. on fused silica (RT)				& STS, Refl.					& 0.36$\pm$0.06		& 2.38$\pm$0.06			& \onlinecite{Rigosi:2016a}\\	
							& CVD on SiO$_2$ (4\,K)			 			& Magneto-refl. 		 	& 0.26\,-\,0.48  & 2.31\,-2\,.53\,***	& \addAlexey{\onlinecite{Stier:2016a}}\\
		\hline
		MoSe$_2$	& MBE on 2L graphene/SiC (5\,K)		& STS, PL	 						& 0.55 						& 2.18				 			& \onlinecite{Ugeda:2014a}\\
							& CVD on HOPG	(79\,K)		 					& STS, PL	 						& 0.5						& 2.15$\pm$0.06			& \onlinecite{Zhang:2015c}\\
		
		\hline
		MoS$_2$		& CVD on HOPG	(77\,K)		 					& STS, PL	 						& 0.2 (or 0.42)						& 2.15$\pm$0.06			& \onlinecite{Zhang:2014d}\\
							& Exf., suspended (77\,K)					& PC			 						& $\geq$0.57 			& 2.5								& \onlinecite{Klots:2014a}\\
							& Exf. on hBN/fused silica (RT)		& PLE									& 0.44$\pm$0.08\,**	& ~2.47$\pm$0.08\,**~	& \onlinecite{Hill:2015a}\\
							& CVD on HOPG (77\,K)			 				& STS, PL 		 				& $\approx$\,0.3\,*	& 2.15$\pm$0.1 			& \onlinecite{chiu:2015}\\
							& Exf. on fused silica (RT)				& STS, Refl.					& 0.31$\pm$0.04		& 2.17$\pm$0.1			& \onlinecite{Rigosi:2016a}\\
  \end{tabular}
	\begin{flushleft}
	\footnotesize
	~\\
	* extracted from the PL data and STS results in Ref. \onlinecite{chiu:2015} \\
	** attributed to the B-exciton transition by the authors of Ref. \onlinecite{Hill:2015a} \\
	*** obtained by adding the estimated exciton binding energies to the resonance energy in Refs. \onlinecite{Stier:2016a,Poellmann:2015} \\
	\end{flushleft}	
  \label{tab:table2}
\end{table*}

\subsubsection{Effective Coulomb potential and the role of the environment }
\label{sec:env} 

Calculations of excitonic states and binding energies in TMD MLs have been performed by many approaches, including effective mass methods, atomistic tight-binding and density functional theory approaches with various levels of sophistication, see, e.g.,~\cite{Cheiwchanchamnangij:2012a,Komsa:2012a,Ramasubramaniam:2012a,Qiu:2013a,Shi:2013a,MolinaSanchez:2013,Bergh:2014,PhysRevB.91.075310,PhysRevB.94.041301,0953-8984-27-34-345003}. 
A simple and illustrative approach to calculate energies of exciton states is provided by the effective mass method. Here, in the Hamiltonian \eqref{HX}, the single-particle kinetic energies $\mathcal H_e(\bm k_e)$ and $\mathcal H_h(\bm k_h)$ are replaced by the operators $-\hbar^2/(2m_e) \partial^2/\partial \bm \rho_e^2$ and $-\hbar^2/(2m_h) \partial^2/\partial \bm \rho_h^2$, respectively, with $\bm \rho_e$, $\bm \rho_h$ being the electron and hole in-plane position vectors. 
Most importantly, the electric field between individual charges in the ML permeates both the material layer and the surroundings outside the monolayer. As a consequence, both the strength and the form of the effective Coulomb interaction between the electron and hole in the exciton are strongly modified by the dielectric properties of the environment \cite{Stier:2016b,Raja:2017a}.
In principle, one recovers a 2D hydrogen-like problem with an adjusted effective potential by taking into account the geometry of the system and the dielectric surroundings~\cite{1979JETPL..29..658K,Cudazzo:2011a,PhysRevB.88.045318,Chernikov:2014a,PhysRevLett.114.107401,Rytova:1967}.

Typically, the combined system  ``vacuum + TMD monolayer + substrate'' is considered, reproducing the main features of the most common experimentally studied samples.
In the effective medium approximation, the dielectric constant $\varepsilon\sim 10$ of the TMD ML generally far exceeds the dielectric constants of the surroundings, i.e., of the substrate $\varepsilon_s$ and of the vacuum. 
As a result, the effective interaction potential takes the form of $\propto 1/\rho$ ($\bm \rho = \bm \rho_e - \bm \rho_h$ is the relative electron-hole coordinate) only at large distances between the particles where the electrical field resides  outside the TMD ML itself. 
At smaller distances, the dependence is $\propto \log(\rho)$~\cite{Cudazzo:2011a}. 
The resulting overall form of the effective potential, following \cite{Rytova:1967,1979JETPL..29..658K}, is approximated by
\begin{equation}
\label{Keldysh}
V(\rho) = -\frac{\pi e^2}{(1+\varepsilon_s)r_0} \left[ \mathbf H_0\left(\frac{\rho}{r_0}\right) - Y_0\left(\frac{\rho}{r_0}\right)\right],
\end{equation} 
where $\mathbf H_0(x)$ and $Y_0(x)$ are the Struve and Neumann functions, $r_0$ is the effective screening length. 
The latter can either be calculated from \textit{ab-intio}~\cite{PhysRevB.88.045318} or considered as a phenomenological parameter of the theory \cite{Chernikov:2014a} and typically ranges from roughly 30~\AA~to 80~\AA.
Then, within the effective mass approximation, the two-particle Schr\"{o}dinger equation with the effective potential $V(\rho)$ in the form of Eq.~\eqref{Keldysh} can be solved, e.g., variationally and numerically or, in some cases analytically~\cite{PhysRevLett.114.107401}.
The result is a series of exciton states described by the envelope functions of the relative motion $\varphi_{nm}(\bm \rho)$. 
Overall, the model potential in the form~\eqref{Keldysh} describes the deviations from the ideal 2D hydrogenic series observed in the experiments and can be used as an input in more sophisticated calculations of excitonic spectra~\cite{Steinhoff:2014,Bergh:2014}.
This simple model potential also agrees well with the predictions from high-level \textit{ab-intio} calculations using Bethe-Salpeter equation approach \cite{Qiu:2013a, Ye:2014a, Ugeda:2014a,Wang:2015b, Latini2015, Chaves2017}.\\
\indent Although a reasonably adequate description of the experimental data for the exciton binding energies is already provided by the relatively simple effective mass model with an effective potential in the form of Eq.~\eqref{Keldysh}, there are several issues debated in the literature that require further studies: \\
$\bullet$ Since the exciton binding energy typically exceeds phonon energies both in TMD ML~\cite{phonons:2d} and in typical substrates, static screening is not necessarily well justified \cite{Stier:2016b}. However, the frequency range at which the screening constant should be evaluated and whether high-energy optical phonons play a role merits further investigation. \\
$\bullet$  Depending on the material and the substrate, the binding energy can be as large as $1/4\ldots 1/3$ of the band gap, see Tab.~\ref{tab:table2}. The excitons have also a relatively small radii leading to a sizable extension of the wavefunction in reciprocal space. Therefore, the effective mass model may not always provide quantitatively accurate results and the effects of the band non-parabolicity and the spin-orbit coupling should be included. \\
$\bullet$  In addition, the trigonal symmetry of the TMD MLs results in the mixing of the excitonic states $(n,m)$ with different $m$  particularly, in the mixing of the $s$- and $p$-shell excitons (i.e., the states with $m=0$ and $|m|=1$) as demonstrated theoretically in Ref.~\cite{Glazov:2017a,Gong:2017a}. Further studies of exciton mixing within ab-initio and tight-binding models to determine quantitatively the strength of this effect are required, in addition to more detailed one and two-photon excitation experiments.\\
$\bullet$ Also the ordering of $2s$ and $2p$ resonances remains an open issue in light of recent theoretical predictions of the state-mixing and the experimental challenges are to precisely determine the $2s/2p$ splitting in TMD MLs, i.e., by comparing the linear and nonlinear absorption spectra or by detailed studies of magneto-optical properties in high-quality MLs, and the eventual splitting of the 2$p$ states \cite{2p:1,PhysRevB.91.075310,Zhouberry:2015a}. \\
$\bullet$ On the experimental side, controlling the influence of the dielectric screening of the surroundings is of particular importance.
Recent works on this topic include observations of exciton states in different solutions \cite{Lin:2014, Ye:2014a}, measurements of changes in the exciton Bohr radii from diamagnetic shifts on different substrates \cite{Stier:2016b} and demonstration of the bandgap and exciton energy renormalization due to external dielectric screening \cite{Raja:2017a}.\\
$\bullet$  Further questions arise with respect to the uniformity of the dielectric environment, with possible variations of the sample-substrate distance and the non-uniform coverage by adsorbates, also considering the recently predicted nanometer spatial sensitivity of the screening effect \cite{Rosner:2016}. Here experimental comparisons between different capped and uncapped samples will be helpful  as well to study, for example, the influence of the substrate morphology on the exciton states. \\
\subsection{Light-matter coupling via excitons}
\label{sec:light}
\subsubsection{Dark and bright excitons}\label{light:dark}

When generated by resonant photon absorption under normal incidence, excitons are optically \textit{bright} (see also discussion in Sec.~\ref{sec:form}).
Subsequent scattering events with other excitons, electrons, or phonons, and defects can induce spin flips and considerable changes in exciton momentum. Alternatively, in case of a more complex generation process such as non-resonant optical excitation or electrical injection, a variety of exciton states can form. As a result of all the above, an exciton may not necessarily be able to recombine radiatively, for instance if the optical transition is now spin forbidden. Such an exciton is described as optically \textit{dark}. Another way to generate \textit{dark} excitons is if a hole and an electron, for instance injected electrically, come together to form an exciton with total angular momentum $\neq1$ or large center-of-mass momentum $\bm K_\text{exc}$. So whether or not excitons can directly interact with light either through the absorption or emission of a single photon, depends on the center of mass wavevector $\bm K_\text{exc}$, the relative motion wavefunction, the valley, $\tau_e$ ($\tau_h$), and spin, $s_e$ ($s_h$), states of the electron and hole. \\
\indent In TMD MLs, exciton-photon coupling is governed by chiral optical selection rules: For normally incident light the direct interband transitions at the $K^\pm$ points of the Brillouin zone are active for $\sigma^\pm$ light polarization, Fig.~\ref{fig:fig1}c,d~\cite{Yao:2008, Xiao:2012a,Zeng:2012a,Cao:2012a,Mak:2012a,Sallen:2012a}. 
Considering interband transitions, the spin and valley states of the electron are conserved and the electron and hole are generated within the same unit cell. 
As a result, the $ns$-shell excitonic states (i.e., those with $m=0$, such as $1s, 2s, 3s$, etc.) where the envelope function $\varphi_{ns}(0)\ne 0$, with $\tau_e=-\tau_h=+1$, $s_e = -s_h=+1/2$ are active in $\sigma^+$ polarization and the states with $\tau_e=-\tau_h=-1$, $s_e=-s_h=-1/2$ are active in $\sigma^-$ polarization. 
The exciton states with $\tau_e=\tau_h$ (occupied electron states in CB and unoccupied electron states in VB) or $s_e=s_h$ (electron and unoccupied state have opposite spins) are dark~\cite{PSSB:PSSB201552211}. 
A schematic illustration of bright and dark electron transitions corresponding to the respective exciton states is presented in Fig.~\ref{fig:fig_light_matter}a.
While the above rules describe the A-exciton series, they are essentially the same for the B-exciton states when the opposite signs of the corresponding spin indices are considered. Also, an admixing of the $p$-character to the $s$-like states is theoretically predicted due to the exchange interaction~\cite{Glazov:2017a,Gong:2017a} and disorder~\cite{Berghauser:2017a} giving rise to modification of selection rules of one- or two-photon transitions.\\
\begin{figure}
\includegraphics[width=0.5\textwidth]{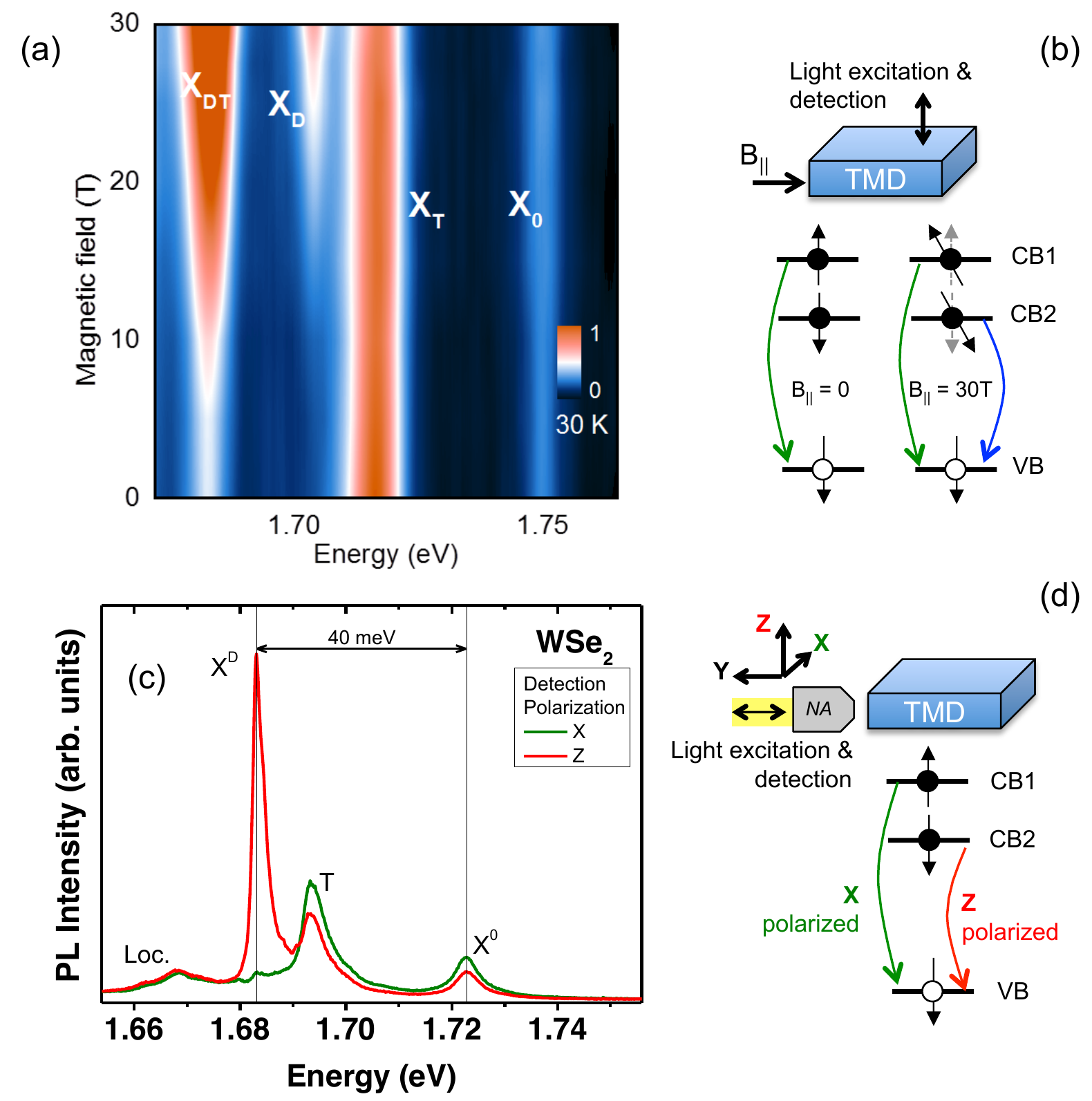}
\caption{\label{FigBrightDark} (a) Brightening of the dark exciton transition observed in ML WSe$_2$ by photoluminescence experiments with in an in-plane magnetic field \cite{Zhang:2017a} (b) Schematic of the brightening of the dark exciton transitions involving the spin states in the conduction band 1 and 2. For simplicity we do not show the Coulomb exchange energy term that also contributes to the dark-bright splitting \cite{Echeverry:2016a}.  (c) and (d) Using in-plane optical excitation and detection, the dark (X$^D$) and bright (X$^0$) exciton can be distinguished by polarization dependent measurements, adapted from \cite{Wang:2017a}. The WSe$_2$ ML is encapsulated in hBN for improved optical quality.}
\end{figure}
\indent For neutral $1s$ excitons, the order and energy difference between bright and dark excitons is given by the sign and amplitude of the spin splitting in the conduction band and the short-range Coulomb exchange interaction, similar to the situation in quantum dots \cite{Crooker:2003a}. 
For WS$_2$ and WSe$_2$, the \emph{electron} spin orientations in the upper valence band and in the lower conduction band are opposite, while in MoS$_2$ and MoSe$_2$, the spins are parallel, as shown in Fig.~\ref{fig:fig1}c,d, although recent studies discuss the possibility of the ground state in ML MoX$_2$ being dark \cite{Molas:2017a,Baranowski:2017a}. 

As a result, the lowest lying CB to VB transition is spin forbidden (dark)  in WS$_2$ and WSe$_2$, the spin allowed transition is at higher energy as indicated in Fig.~\ref{FigBrightDark}. 
One experimental approach to measure the energy splitting between the dark and bright state is to apply a strong in-plane magnetic field. This leads to an admixture of bright and dark states which allows detection the dark transitions that gain oscillator strength and appear in the spectrum as the magnetic field increases, see Fig.~\ref{FigBrightDark}a,b \cite{Zhang:2017a,Molas:2017a}. For ML WSe$_2$, the dark excitons lie about 40~meV below the bright transitions.
In addition to spin conservation, there is another important differences between the so called bright and dark excitons: Symmetry analysis \cite{Glazov:2014a,Slobodeniuk:2016a,Wang:2017b,Zhou:2017a} shows that the spin-forbidden dark excitons are optically allowed  with a dipole out of the monolayer plane ($z$-mode), whereas the spin-allowed bright excitons have their dipole in the monolayer plane $xy$. Therefore optical excitation and detection in the plane of the monolayer (i.e., in the limit of grazing incidence) allows a more efficient detection of these in principle spin-forbidden transitions than experiments with excitation/detection normal to the monolayer, as indicated in Fig.~\ref{FigBrightDark}c,d. This $z$-mode exciton transition can be clearly identified by its polarization perpendicular to the surface using a linear polarizer. Another approach is to couple the $z$-mode to surface plasmons for polarization selectivity as in  \textcite{Zhou:2017a}. Using these techniques, the same dark-bright exciton splitting as reported in the magnetic-field dependent experiments, namely 40-50 meV, could be extracted for ML WSe$_2$.
The origin of the $z$-mode transition, which remains very weak compared to the spin-allowed exciton, lies in mixing of bands with different spin configuration and orbital origin.\\
\indent Of similar origin as the spin-forbidden \emph{intra-}valley dark excitons are the spin-allowed \emph{inter-}valley states, where the direct transition of the electron from the valence to conduction band is forbidden due to the momentum conservation.
Examples are inter-valley $K^{\pm}$-$K^{\mp}$, $K^\pm$-$Q$, $\Gamma$-$K^\pm$ and $\Gamma$-$Q$ excitons, where $K^\pm$, $Q$ and $\Gamma$ refer to the particular points in the Brillouin zone.

\begin{figure}[ht!]
\includegraphics[width=0.48\textwidth]{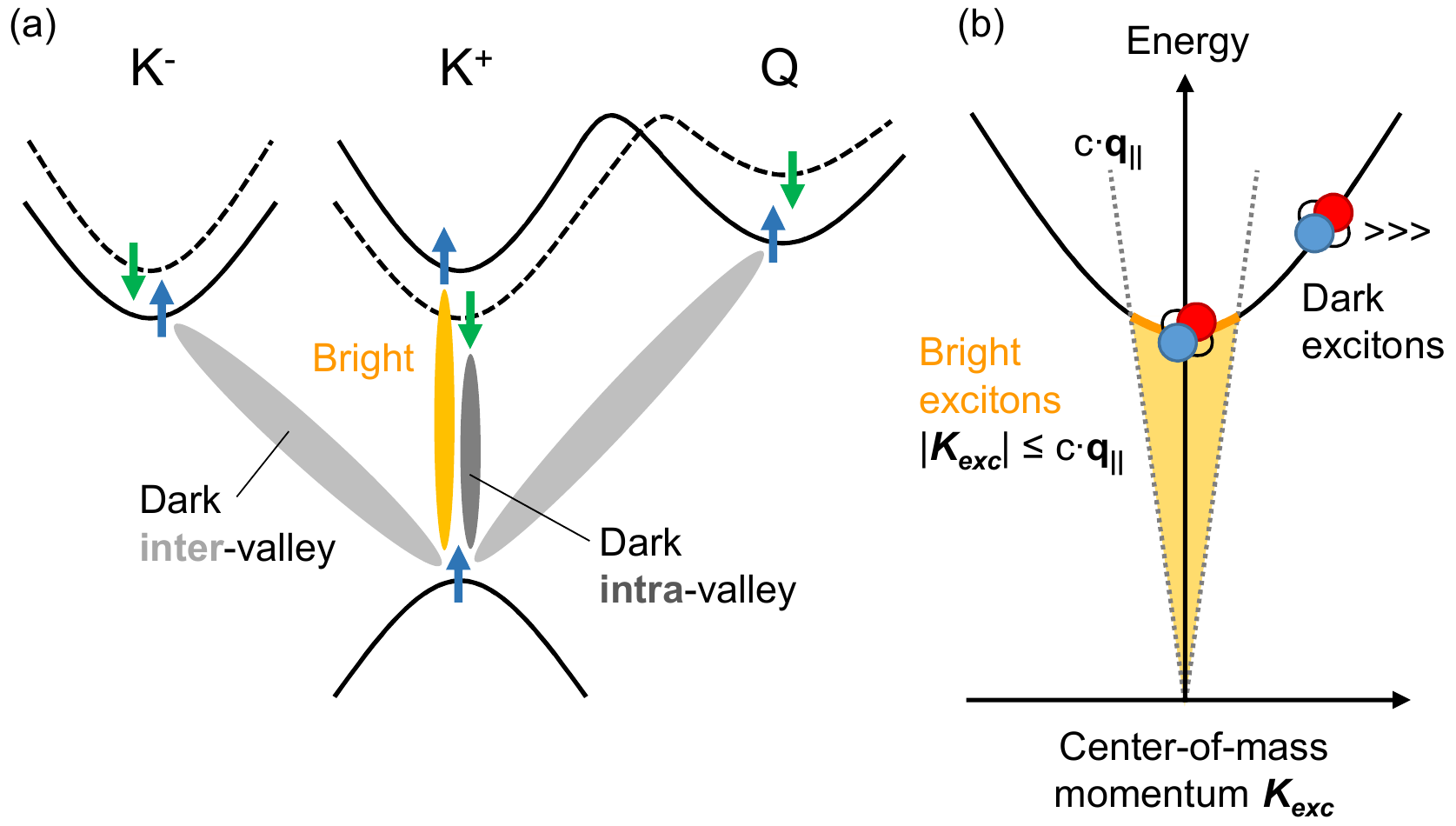}
\caption{\label{fig:fig_light_matter} (a) A schematic overview of typical allowed and forbidden electronic transitions for the respective bright and dark exciton states. 
The underlying band structure is simplified for clarity, including only the upper valence band at $K^+$ and the high-symmetry points $K^\pm$ and $Q$ in the conduction band. The order of the spin states in the conduction band, corresponds to W-based TMD MLs, see Refs.~\cite{Liu:2013a, Glazov:2014a, Kormanyos:2015a} for details.
(b)~Schematic illustration of the exciton ground-state dispersion in the two-particle representation.
The light-cone for bright excitons is marked by the free-space photon-dispersion, $c q_\parallel$,  where $c$ is the speed of light and the excitons outside of the cone are essentially dark.
}
\end{figure}

\subsubsection{Radiative lifetime}
\label{sec:radlife}

An additional constraint on the optical activity of the excitons is imposed by the center-of-mass wavevector conservation $\bm K_\text{exc}$, which should be equal to the projection of the photon wavevector $\bm q_\parallel$ on the TMD ML plane.
The range of the wavevectors meeting this requirement obeys, for a ML in vacuum, the condition $K_\text{exc}\leqslant q_0=\omega_0/c$, where $\omega_0$ is the photon frequency corresponding to the exciton resonance (for $1s$ exciton $\omega_0 = (E_g - E_B)/\hbar$).
Bright excitons within this so-called ``light cone'' couple directly to light, i.e., can be either be created by the absorption of a photon or spontaneously decay through photon emission, while excitons with $K_\text{exc}>q_0$ are optically inactive. \\
\indent In general, the radiative decay rate $\Gamma_0$ of the bright excitons within the light cone, which also determines the overall strength of optical absorption (i.e., total area of the resonance), is proportional to the probability of finding the electron and the hole within the same unit cell, i.e., to $|\varphi_{ns}(0)|^2\propto 1/a_B^2$, where $a_B$ is the effective Bohr radius.  
The strong Coulomb interaction in TMD MLs, leading to the large binding energies of the excitons, also results in relatively small exciton Bohr radii, $a_B\sim 1$~nm for the $1s$ state, as discussed above. Estimates of $\Gamma_0$ for the $1s$ exciton within a simple two-band model~\cite{Glazov:2014a,PSSB:PSSB201552211} then yield $\hbar\Gamma_0 \gtrsim 1$~meV. This corresponds to a radiative decay time $1/(2\Gamma_0) \lesssim 1$~ps, in good agreement with experimental observations \cite{Poellmann:2015,Moody:2015,Palummo:2015,Jakubczyk:2016a,Robert:2016a}. 
Hence, the radiative decay times of excitons in TMD MLs are about two orders of magnitude shorter as compared, e.g., with the excitons in GaAs-based quantum wells \cite{Deveaud:1991a}.
In addition, the radiative broadening on the order of 1\,meV imposes a lower limit on the total linewidth of the bright exciton resonance \cite{Cadiz:2017a,Moody:2015,Jakubczyk:2016a,Dey:2016a}.
This simple analysis is further corroborated by first principle calculations, which predict exciton intrinsic lifetimes as short as hundreds of fs \cite{Palummo:2015,Wang:2016a}.\\
\indent Importantly, the presence of the radiative cone determines the overall effective decay rate of an exciton population at finite temperatures through the radiative recombination channel. Which fraction of excitons is within and which fraction is outside the light cone depends on temperature ~\cite{Andreani:1991a}. The \emph{effective} radiative decay for thermalized populations is obtained from the radiative decay rate within the light cone $\Gamma_0$, weighted by the fraction of the excitons inside the cone.
In case of strictly 2D systems with a parabolic exciton dispersion, above very low temperatures, this fraction decreases linearly with the temperature~\cite{Andreani:1991a}.
For MoS$_2$, the effective radiative recombination time is calculated to be on the order of several tens of ps at cryogenic temperatures and to exceed a nanosecond at room temperature~\cite{Wang:2016a}. While radiative recombination is forbidden outside the light cone if wavevector conservation holds, this can be partially relaxed due to the presence of disorder caused, e.g., by impurities or defects, since momentum conservation is relaxed in disordered systems~\cite{PhysRevB.47.3832,Vinattiery}.\\
\indent The effective radiative lifetime is, of course, also affected by the presence of the spin-forbidden intra-valley and inter-valley dark states considering thermal distribution of excitons between these states.  It further depends on the relaxation rate of the dark excitons of the reservoir towards low-momentum states \cite{PhysRevB.94.205423}, potentially leading to the additional depletion of the excitons within the radiative cone~\cite{Kira:2005}.
When the excitons are predominantly created within the radiative cone through resonant or near-resonant excitation, an initial ultra-fast decay has been indeed observed\,\cite{Poellmann:2015, Robert:2016a} and attributed to the intrinsic radiative recombination time $\Gamma_0$ of the bright states.
The excitons were shown to thermalize subsequently and to experience slower decay at later times. At room temperature, effective radiative exciton lifetimes as long as 20~ns have been measured in super-acid treated samples \cite{amani:2015a} and estimated to be on the order of 100~ns from combined time-resolved PL and quantum yield measurements \cite{Jin:2016b}.\\
\indent Finally we note, that the overall decay of the exciton population is usually governed by the complex interplay of radiative and non-radiative channels.
It is thus affected by the presence of defects and disorder, Auger-type exciton-exciton annihilation at elevated densities~\cite{Kumar:2014b,Mouri:2014a,Sun:2014a,Yu:2015b}, and through the formation of exciton complexes such as biexcitons~\cite{You:2015a,Sie:2015a} and trions~\cite{Mak:2013a,Ross:2013a}. 
Finally, radiative recombination itself depends on the optical environment, i.e., the effective density of the photon modes available as final states for the recombination of the excitons.
The effective strength of the light-matter interaction is thus modified by the optical properties of the surroundings (e.g., refractive index of the substrate) and can be tuned externally.
The integration of the TMD MLs in optical cavities highlights this aspect. Indeed, the strong-coupling regime has been demonstrated, where excitons and photons mix to create hybrid quasiparticles, exciton polaritons~\cite{Liu:2015b,Dufferwiel:2015a,Vasilevskiy:2015a,Lundt:2016a,Sidler:2016,Flatten:2016a}. The discussion above highlights the complex challenges for interpreting for example photoluminescence emission times measured in experiments in terms of intrinsic decay rates, effective radiative lifetimes and non-radiative channels, for example.

\subsubsection{Exciton formation}
\label{sec:form}
In many spectroscopy experiments performed on TMDs monolayers involving optical injection, the excitation laser energy is larger than the exciton ground state energy. This means that exciton formation dynamics and energy relaxation have to be taken into account. Two exciton formation processes are usually considered in semiconductors: (i) direct hot exciton photogeneration, with the simultaneous emission of phonons, in which the constitutive electron-hole pair is geminate \cite{Bonnot:1974a}; or (ii) bimolecular exciton formation which consists of direct binding of electrons and holes  \cite{Barrau:1973a,Piermarocchi:1997a}. In 2D semiconductors based on GaAs quantum wells the bimolecular formation process plays an important role \cite{Amand:1994a,Piermarocchi:1997a,Szczytko:2004a}.
When the excitation energy lies below the free particle bandgap in TMD monolayers, the exciton formation process can only be geminate (neglecting Auger like and two-photon absorption effects). Note that this process, which involves a simultaneous emission of phonons, can yield the formation of either intra-valley or inter-valley excitons.
When the excitation energy is strongly non-resonant, i.e. above the free particle bandgap, the PL dynamics is very similar compared to the quasi-resonant excitation conditions in MoS$_2$ or WSe$_2$ monolayers \cite{Korn:2011a,Wang:2014b, Zhang:2015d}. The PL rise time is still very short and no signature of bimolecular formation and energy relaxation of hot excitons can be evidenced, in contrast to III-V or II-VI quantum wells. Indeed, recent reports indicate ultra-fast exciton formation on sub-ps timescales after non-resonant excitation \cite{Cha:2016,Ceballos:2016,Steinleitner:2017}. While further studies are required, at this stage one can already speculate that the strong-exciton phonon coupling in TMD monolayers seems to yield efficient exciton formation process for a wide range of excitation conditions. We also note, that alternative processes such as multi-exciton generation, i.e., the reverse of Auger-type annihilation, might become important for sufficiently high excess energies.

\section{Excitons at finite carrier densities}
\label{sec:complex}

The discussion in the previous Section \ref{sec:Coulomb} deals with the fundamental properties of the excitons in TMD MLs in the low-density regime.
However, the presence of photoexcited carriers, either in the form of Coulomb-bound or free charges, can significantly affect the properties of the excitonic states, as is the case for traditional 2D systems with translational symmetry, such as quantum wells~\cite{Haug2009}.

\subsection{The intermediate and high density regimes}
We distinguish two partially overlapping regimes of \emph{intermediate} and \emph{high} density conditions.
These can be defined as follows: In the intermediate density regime the excitons can still be considered as bound electron-hole pairs, but with properties considerably modified compared with the low-density limit.
In the high density regime, beyond the so-called Mott transition, excitons are no longer bound states; the electrons and holes are more appropriately described as a dense Coulomb-correlated gas. Under such conditions, the conductivity of the photoexcited material behaves less like the insulating semiconductor with neutral excitons and more like a metal with many free carriers, whence the description of this effect as a photoinduced Mott transition. 
The transition between two regimes is controlled by the ratio of the average carrier-carrier (or, alternatively, exciton-exciton) separation $2/\sqrt{n\pi}$ to the exciton Bohr radius $a_B$ at low density: For $2/(\sqrt{n\pi}a_B) \lesssim 1$ the density of carriers (or excitons) $n$ can be considered as high.
Due to the small Bohr radius of about 1\,nm in TMD MLs, the intermediate and high density regimes are reached at significantly higher carrier densities compared to systems with weaker Coulomb interactions, such as III-V or II-VI semiconductor quantum wells. 
With respect to absolute numbers, the intermediate case with inter-particle distances about 100 to $10 \times a_B$, broadly covers the density range between $10^{10}$ and several $10^{12}$\,cm$^{-2}$.
The high density case then corresponds to separations on the order of a few Bohr radii or less and is considered to apply for carrier densities of a few $10^{13}$ to $10^{14}$\,cm$^{-2}$ or higher.
In particular, the electron-hole pair-density of $n=a_B^{-2}$, often used as a rough upper estimate for the Mott transition \cite{Klingshirn:2007}, yields $n \sim 10^{14}$\,cm$^{-2}$ for TMD MLs.

The main phenomena occurring at elevated \emph{carrier} densities can be briefly summarized as follows: \\
$\bullet$ \textbf{First}, there are efficient scattering events.
Elastic and inelastic scattering of excitons with free carriers or excitons leads to relaxation of the exciton phase, energy, momentum and spin and thus to spectral broadening of the exciton resonances~\cite{Wang:1993,Shi:2013b,Moody:2015,Chernikov:2015a,Dey:2016a}. 
In addition, through inelastic scattering with free charge carriers, an exciton can capture an additional charge and form a bound three-particle state at intermediate densities, the so-called \emph{trion} states ~\cite{Stebe:1989aa,Kheng:1993,Mak:2013a,Ross:2013a,Singh:2016a}. 
Similarly, at intermediate \emph{exciton} densities, interactions between excitons can result in a bound two-exciton state, the \emph{biexciton} state ~\cite{PhysRevB.25.6545,You:2015a,Sie:2015a,Plechinger:2015a,Shang:2015}, resembling the hydrogen molecule. 

Charged excitons (trions) and biexcitons were predicted for bulk semiconductors~\cite{PhysRevLett.1.450} by analogy with molecules and ions.
While they naturally appear as a result of Coulomb interactions between three or four charge carriers, we also note that in real systems with finite carrier densities, the correlations between, e.g., excitons/trions and the Fermi sea of electrons (or holes) may be of importance~\cite{Suris:2003,Sidler:2016,Efimkin:2017a}. 
Furthermore, excitons formed from two fermions can be considered as composite bosons at least for not too high carrier densities. Interestingly, excitons are expected to demonstrate at low to intermediate densities collective phenomena such as Bose-Einstein condensation (strictly speaking, quasi-condensation in two-dimensions) and superfluidity~\cite{Moskalenko62:eng,keldysh68a,Fogler:2014a}. First signatures of boson scattering of excitons in monolayer WSe$_2$ have been reported \cite{Manca:2017a}. 
Additionally, exciton-exciton scattering can also lead to an Auger-like process: the non-radiative recombination of one exciton and dissociation of the other into an unbound electron and hole, leading to exciton-exciton annihilation, as already mentioned in Sec.~\ref{sec:radlife} ~\cite{Kumar:2014b,Mouri:2014a,Sun:2014a,Yu:2015b,Robert:2016a}.\\
$\bullet$ \textbf{Second}, finite quasiparticle densities generally lead to what can be broadly called dynamic screening of the Coulomb interaction~\cite{Haug2009, Klingshirn:2007}.
In analogy to the behavior of quasi-free carriers in metals, it is related to both direct and exchange contributions and typically decreases the effective strength of the Coulomb interaction.
As a result of the decreasing electron-hole attraction, the exciton binding energy is reduced; the average electron-hole separation increases, thus also leading to lower oscillator strengths for excitonic transitons i.e. a to weaker light-matter coupling. 
In addition, the photoinduced screening induces renormalization of the free particle band gap to lower energies.
In many cases, including TMD MLs, the decrease of the exciton ground-state ($n=1$) binding energy and the red shift of the bandgap are of similar magnitude, at least in the intermediate-density regime.
Hence, while the absolute shifts of the $n=1$ resonance, i.e., of the optical band gap (see Fig.~\ref{fig:fig2}), can be rather small, on the order of several tens of meV, the underlying changes in the nature of excitations (binding energies, free-particle band gap) are about an order of magnitude larger~\cite{Steinhoff:2014, Chernikov:2015a, Chernikov:2015b, Ulstrup:2016}.\\
$\bullet$ \textbf{Third}, the presence of free carriers decreases the available phase space for the electron-hole complexes due to the Pauli blocking~\cite{Haug2009}. 
This also results in a decrease of trion and exciton binding energies and the oscillator strengths. In addition, at sufficiently high densities of both electrons and holes, it results in population inversion, i.e., more electrons populating the conduction rather than valence band over a certain range of energy. As in quantum wells \cite{Haug2009}, this regime is expected to roughly coincide with the Mott transition discussed above. Moreover, in the high-density regime, bound electron-hole states cannot be formed and thus the optical spectra are no longer dominated by the exciton resonance. Population inversion then leads to stimulated emission processes and negative absorption for the corresponding transitions ~\cite{Haug2009, Chernikov:2015b}.
In the absence of competing scattering and absorption channels in the respective energy range, this would give rise to amplification of radiation and allow in principle for the use of the material as an active medium in lasing applications; see Refs.~\cite{Wu:2015b, Ye:2015a, Salehzadeh:2015} for reports of lasing in TMD MLs. \\
\indent Many issues in the high-density regime still remains to be explored, both experimentally and theoretically, the prepondence of literature on TMD monolayers having addressed the behavior of the materials at intermediate densities~\cite{Korn:2011a,Wang:2013d,Lagarde:2014a,Singh:2014a,Mai:2014a,Kumar:2014a,Zhucr:2014a,Poellmann:2015, Schmidt:2016a1}. We also note that an accurate, quantitative treatment of many-body physics of strongly interacting systems is a very challenging problem. Promising steps in that direction are presented, for example in ~\cite{Steinhoff:2014, Steinhoff:2015, Schmidt:2016a1, Selig:2016}. Although effects related with occupation of other ($Q$ and $\Gamma$) valleys with an increase in the free-carrier density are of interest, the relative simplicity of the electronic structure of TMD monolayers, their tunability under external conditions and dielectric media, and experimental accessibility and their strong many-body effects make these systems promising test cases for advancing our understanding of fundamental issues in many-body interactions at high densities. 


\subsection{Electric charge control}
\label{sec:charge}

\begin{figure}
\includegraphics[width=0.45\textwidth]{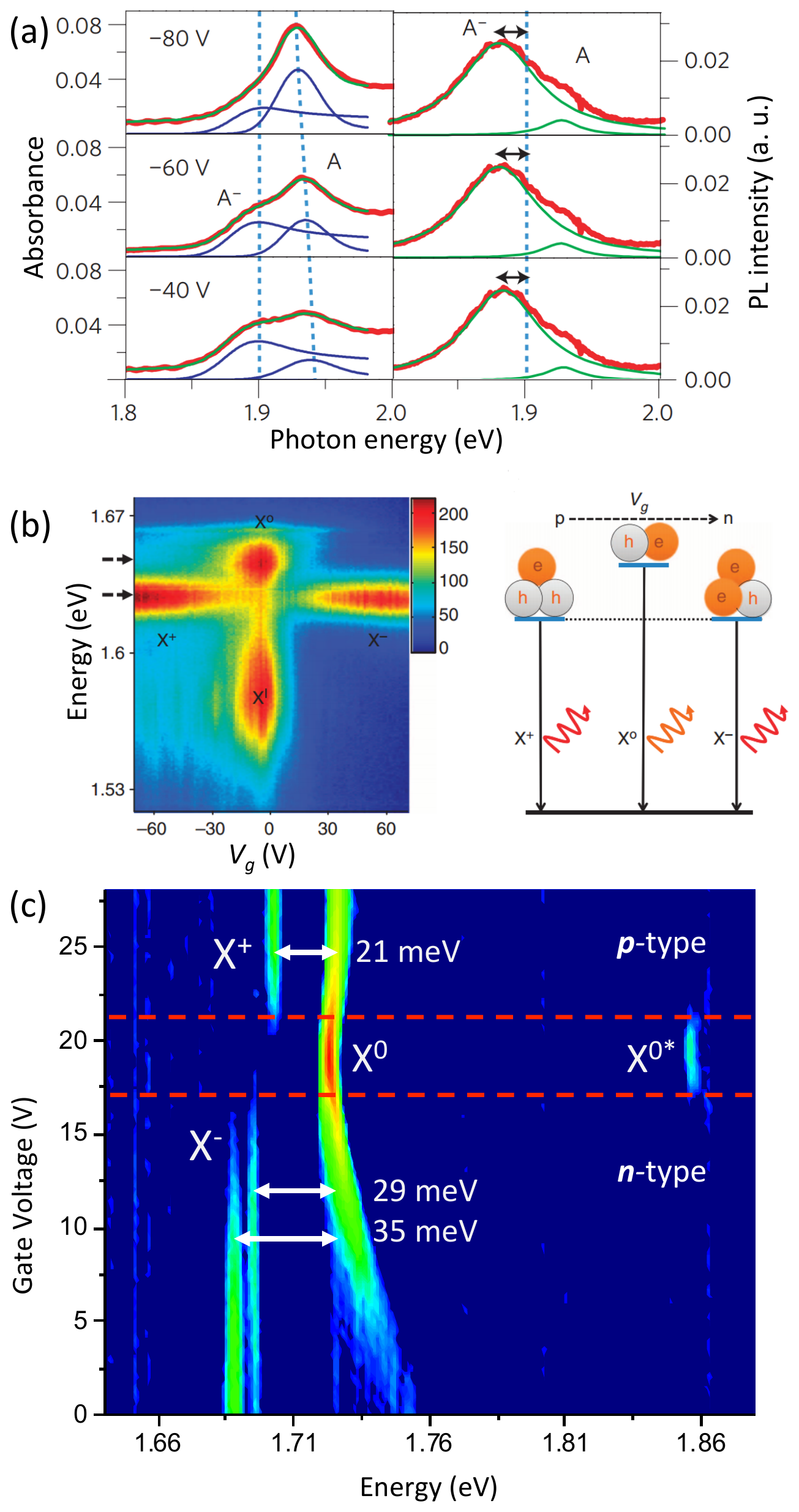}
\caption{\label{fig:chargetune} (a) Absorbance and photoluminescence experiments exhibiting signatures of neutral (A) and charged (A$^-$) excitons in a charge tunable \textbf{MoS$_2$} monolayer \cite{Mak:2013a}. (b) Color contour plot of PL from an electrically gated \textbf{MoSe$_2$} monolayer that can be tuned to show emission from positively charged (X$^+$) to negatively charged (X$^-$) trion species \cite{Ross:2014a}. (c) Contour plot of the first derivative of the differential reflectivity in a charge tunable \textbf{WSe$_2$} monolayer. The $n$- and $p$-type regimes are manifested by the presence of X$^+$ and X$^-$ transitions. Around charge neutrality, the neutral exciton X$^0$ and an excited state X$^{0*}$ are visible \cite{Courtade:2017a}.
}
\end{figure}

While neutral excitons tend to dominate the optical properties of ML TMDs, more complex exciton species also play an important role. Particularly prevalent are charged excitons or  \emph{trions}, the species formed when an exciton can bind another electron (or hole) to form a negatively (or positively) charged three-particle state. Since unintentional doping in TMD layers is often $n$-type \cite{Radisavljevic:2011a,Anthony:2007}, the formation of negative trions is likely, assuming that adsorbates do not introduce additional significant changes to the doping level. In general, the trion binding energy in semiconductor nano-structures is typically 10\% of the exciton binding energy \cite{Donck:2017a}.
For a neutral exciton binding energy on the order of 500~meV, this yields an estimated trion binding energy of several tens of meV. \\
\indent In monolayer MoS$_2$, Mak and coworkers observed tightly bound negative trions with a binding energy of about 20~meV \cite{Mak:2013a}, see Fig.~\ref{fig:chargetune}a, which is one order of magnitude larger than the binding energy in well-studied quasi-2D systems such as II-VI quantum wells \cite{Kheng:1993}, where trions were first observed. At low temperature in monolayer MoSe$_2$, well-separated neutral and charged excitons are observed with a trion binding energy of approximately 30~meV, as clearly demonstrated in charge tunable structures \cite{Ross:2013a}, see Fig.~\ref{fig:chargetune}b. 
In this work, the authors also show the full bipolar transition from the neutral exciton to either positively or negatively charged trions, depending on the sign of the applied gate voltage. 
The binding energies of these two kinds of trion species were found  to be similar, an observation  consistent with only minor differences in the  effective masses of electrons and holes in most of the studied TMDs \cite{Liu:2013a,Kormanyos:2015a}. We also note, that in optical spectra, the energy separation between neutral excitons and trions is a sum of the trion binding energy (strictly defined for the zero-density case) and a term proportional to the Fermi energy of the free charge carriers, see, e.g., \cite{Mak:2013a,Chernikov:2015a}. In addition to the trion signatures in PL and at sufficiently large free carrier densities, the signatures of the trions are also found in absorption-type measurements \cite{Mak:2013a,Jones:2013a,Chernikov:2014a,Chernikov:2015a,Singh:2016b}. \\
\indent Electrical charge tuning of excitons is commonly observed in monolayer TMDs devices, also including WSe$_2$ \cite{Jones:2013a} and WS$_2$ \cite{Plechinger:2015a,Shang:2015}. 
In WS$_2$, these two works also reported biexcitons in addition to neutral and charged excitons.\\ 
\indent As a fundamental difference to conventional quantum well structures, in monolayer TMDs the carriers have an additional degree of freedom: the valley index.
This leads to several optically bright and also dark configurations, for a classification, see e.g. \cite{Yu:2015v,Dery:2015a,PhysRevLett.114.107401,Courtade:2017a}, which can give rise to potentially complex recombination and polarization dynamics \cite{Volmer:2017a}. Charge tunable monolayers that are encapsulated hexagonal boron nitride, result in narrow optical transitions, with low-temperature linewidths typically below 5~meV, as shown in Fig.~\ref{fig:chargetune}c. This has revealed the trion fine structure related to the occupation of the same or different valleys by the two electrons \cite{Jones:2013a,Jones:2016aa,Plechinger2016,Singh:2016b,Courtade:2017a}. 
An informative comparison between charge tuning in ML WSe$_2$ and ML MoSe$_2$ was recently reported in \textcite{Wang:2017a} and revealed the highest-energy valence band and the lowest-energy conduction band to have antiparallel spins in ML WSe$_2$, and parallel spins in ML MoSe$_2$. The concept of the trion as a three particle complex is useful at low carrier densities; at elevated densities intriguing new many-body effects have been predicted by several groups \cite{Dery:2016a,Efimkin:2017a,Sidler:2016}. \\

\section{Valley polarization dynamics}
\label{sec:valley}

\subsection{Valley-polarized excitons}
\label{sec:depol}

Optical control of valley polarization is one of the most fascinating properties of TMD monolayers. In the majority of cases, due to the strong Coulomb interaction, the valley dynamics of photogenerated electrons and holes cannot be adequately described within a single-particle picture as excitonic effects also impact the polarization dynamics of the optical transitions. As previously discussed and predicted in Refs. \cite{Xiao:2012a,Cao:2012a}, optical valley initialization is based on chiral selection rules for interband transitions: $\sigma^+$ polarized excitation results in the inter-band transitions in the $K^+$ valley, and, correspondingly, $\sigma^-$ polarized excitation results in transitions in the $K^-$ valley. Initial experimental confirmation of this effect was reported in steady-state PL measurements in  MoS$_2$ monolayers \cite{Zeng:2012a,Mak:2012a,Cao:2012a,Sallen:2012a}, as well as in WSe$_2$ and WS$_2$ systems \cite{Jones:2013a,Wang:2014b,Mai:2014b,Sie:2015b,Kim:2014z,Zhu:2014b}. The overall degree of polarization has been shown to reach almost unity. In ML MoSe$_2$, however, non-resonant excitation usually results in at most 5\% PL polarization \cite{Wang:2015a}, the reason for this difference remaining a topic of ongoing discussion. Interestingly, for MoSe$_2$, the application of a {strong out-of-plane} magnetic field {combined with} resonant {or nearly resonant} optical excitation appears to be necessary to initialize large valley polarization~\cite{Kioseoglou:2016a}. Finally, in addition to optical valley initialization, strong circularly polarized emission is also reported from electro-luminescence in TMD-based light-emitting devices -- an interesting and technologically promising observation \cite{zhang:2014e,Onga:2016,Yang:2016}. 

\begin{figure*}
\includegraphics[width=0.77\textwidth]{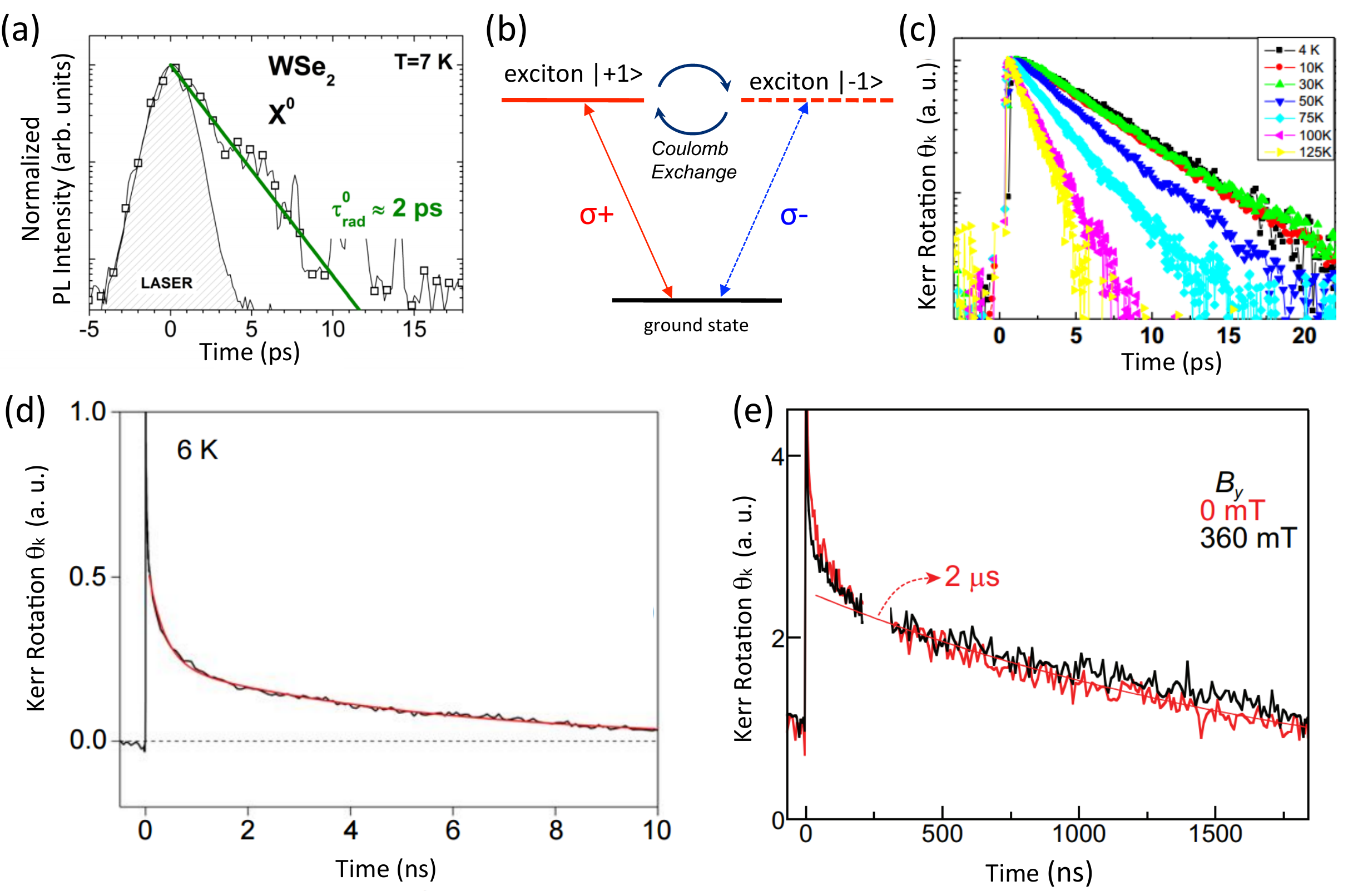}
\caption{\label{fig:fig7} (a) Exciton PL emission time of the order of 2~ps measured in time-resolved photo-luminescence for ML WSe$_2$ at $T=7$~K \cite{Robert:2016a}. (b) Schematic showing that $|+1\rangle$ and $|-1\rangle$ neutral excitons are coupled by the electron-hole Coulomb exchange interaction \cite{Glazov:2014a}. (c)  Decay of the \textbf{neutral exciton} polarization in WSe$_2$ monolayers on ps time scales as measured by Kerr rotation \cite{Zhucr:2014a} (d) Decay of resident \textbf{electron} polarization as measured by Kerr rotation in monolayer WS$_2$, with a typical time constant of 5~ns \cite{Bushong:2016a} for $T=6$~K. (e) Decay of \textbf{hole} polarization in a charge tunable WSe$_2$ monolayer with a time constant of 2$\mu$s \cite{Dey:2017a} for $T=4$~K, where $B_y$ is the magnetic field applied in the sample plane.
}
\end{figure*}

As previously discussed in Sec.~\ref{sec:Coulomb}, following excitation with 
circularly polarized light across the band gap, an exciton is formed from carriers in a 
specific $K$ valley due to the robust, valley dependent optical selection rules. The 
degree of circular polarization $P_c$, as measured in steady state PL, can be 
approximated as $P_c=P_0/(1+\tau/\tau_s)$, where $\tau$ is 
exciton lifetime, $\tau_s$ is the polarization lifetime and $P_0$ is the initially generated polarization. High $P_c$ in steady state PL experiments generally results 
from a specific ratio of $\tau$ versus $\tau_s $ and does not necessary require particularly long polarization lifetimes. Hence the extrinsic parameters such as, e.g., short carrier lifetimes due to non-radiative channels can strongly affect this value and detailed analysis of steady-state experiments is challenging.\\
\indent Time-resolved studies provide more direct access to the valley dynamics of excitons.
In particular, the determination of the exciton PL emission times on  the order of several to tens of picoseconds in typical samples at low temperature, together with measurements of the polarization dynamics indicate that the neutral exciton looses its initial valley polarization very quickly, over a few ps. 
This observation is difficult to understand at the level of individual electrons and holes: The valley polarization in TMDs monolayers should be very stable from within single-particle picture as it requires inter-valley scattering with change in momentum, typically combined with additional electron and hole spin flipping  \cite{Xiao:2012a}. Spin conserving inter-valley scattering is generally energetically unfavorable due to spin splittings of several hundreds and tens of meV in the valence and conduction bands, respectively \cite{Kormanyos:2015a}. In considering the valley dynamics following optical excitation, it is, however, crucial to note that rather than observing individual spin and valley polarized carriers, we create and probe the dynamics of valley-polarized excitons.\\
\indent The Coulomb interaction between the charge carriers does, in fact, strongly impact the valley dynamics in TMD MLs: The long-range exchange interaction between the electron and hole forming an exciton gives rise to a new and efficient decay mechanism for the exciton polarization~\cite{yu:2014e,Glazov:2014a,Yu:2014a,Hao:2016a,Zhucr:2014a}. Indeed, the $\bm k\cdot \bm p$-interaction results in the admixture of the valence band states in the conduction electron state and of the conduction band states in the hole state in the exciton. As a result of this admixture and of the Coulomb interaction, an exciton with an electron in the $K^+$ valley can effectively recombine and produce to an exciton with an electron in the $K^-$ valley. This process needs neither the transfer of significant momentum of an individual carrier nor its spin flip. It can be interpreted in a purely electrodynamical way if one considers an optically active exciton as a microscopic dipole oscillating at its resonant frequency. Naturally, this mechanism is efficient only for bright exciton states and the dark states are largely unaffected. For a bright exciton propagating in the ML plane with the center of mass wavevector $\bm K_{\rm exc}$, the proper eigenstates are the linear combinations of states active in the $\sigma^+$ and $\sigma^-$ circular polarization: One eigenstate has a microscopic dipole moment oscillating along the wavevector $\bm K_{\rm exc}$, this is the longitudinal exciton, and the other one has the dipole moment oscillating perpendicular to the $\bm K_{\rm exc}$, being the transverse exciton. The splitting between those states, i.e., the longitudinal-transverse splitting, acts as an effective magnetic field and mixes the $\sigma^+$ and $\sigma^-$ polarized excitons, which are no longer eigenstates of the system, leading to depolarization of excitons~\cite{Maialle:1993a,Glazov:2014a,Ivchenko:2005a,PSSB:PSSB201552211}. As compared with other 2D excitons, e.g., in GaAs or CdTe quantum wells, in TMD MLs the longitudinal-transverse splitting is enhanced by one to two orders of magnitude due to the tighter binding of the electron to the hole in the exciton and, correspondingly, the much higher oscillator strength of the optical transitions \cite{Li:2014b}. This enhanced longitudinal-transverse splitting as compared to GaAs quantum wells leads to a comparatively faster exciton polarization relaxation. This mechanism, here discussed in the context of valley polarization, also limits valley coherence times \cite{Glazov:2014a,Hao:2016a}, see below. \\
\indent Experimentally, the valley polarization dynamics can be monitored by polarization-resolved time-resolved photoluminescence (TRPL) and pump-probe measurements. By using time-resolved Kerr rotation, Zhu $et~al.$ found that in monolayer WSe$_2$ the exciton valley depolarization time is around 6~ps at 4K, in good agreement with the Coulomb exchange mediated valley depolarization \cite{Zhucr:2014a,Yan:2015b}, see Fig.~\ref{fig:fig7}c. In ML MoS$_2$ and MoSe$_2$ fast exciton depolarization times ($\approx~$ps) were also reported \cite{Lagarde:2014a,Mai:2014a,Wang:2013d,Jakubczyk:2016a}. All these experiments demonstrate measurable depolarization of the neutral exciton X$^0$, although the exact relaxation time may be different in specific measurements depending on the samples used, experimental conditions and techniques employed. \\
\indent Valley depolarization due to the long-range Coulomb exchange is expected to be less efficient for spatially indirect excitons, where the electron-hole overlap is weaker. This configuration applies to type II ML TMD heterostructures, where holes reside in WSe$_2$ and electrons in MoSe$_2$, for example. Indeed \textcite{Rivera:2015a,Rivera:2016a} have observed valley lifetimes of tens of ns for indirect excitons at low temperature, which motivates further valley dynamics experiments in structures with tunable Coulomb interactions, albeit with more complex polarization selection rules. Another type of excitons that is, in principle, unaffected by valley depolarization through Coulomb exchange are optically \textbf{dark} excitons. {With a slight mixing of bright excitons with dark excitons (for optical readout), the dark excitons may provide a promising alternative configuration for exciton valley manipulation \cite{Zhang:2017a}. \\
\indent In addition to the role of Coulomb exchange effects on valley polarization, other mechanisms linked to disorder in the sample and the associated scattering with impurities and phonons have also been investigated, with further details in \cite{Neumann:2017a,Tran:2017a,McCreary:2017a,Yu:2016a}.

\subsection{Valley coherence}
\label{sec:valcoh}

As discussed in the previous section, excitation with circularly polarized light can induce valley polarization in a TMD monolayer  \cite{Xiao:2012a}. Similarly, excitation with linearly polarized light can generate valley coherence, i.e., a \textit{coherent} superposition of $K^+$ and $K^-$ valley states, as first reported for the neutral exciton in ML WSe$_2$ \cite{Jones:2013a}. A fingerprint of generated valley coherence is the emission of \textit{linearly} polarized light from the neutral exciton, polarized along the same axis as the polarization of the excitation, an effect also termed optical alignment of excitons in the earlier literature \cite{Meier:1984a}. In addition, valley coherence in the ML is sufficiently robust to allow rotation of the coherent superposition of valley states in applied magnetic fields \cite{Wang:2016b,Schmidt:2016a,Cadiz:2017a} or with the help of a pseudo-magnetic field generated by circularly polarized light via the optical Stark effect \cite{Ye:2017a}. 

\subsection{Valley polarization dynamics of trions and free charge carriers}
\label{sec:polcomplex}

For manipulating valley polarization of bright, direct excitons within the radiative cone, the radiative lifetime in the ps range sets an upper bound for the available time scale. In addition neutral exciton valley polarization of the neutral exciton decays rapidly due to the Coulomb-exchange mediated mechanism discussed above and shown in Fig.~\ref{fig:fig7}c. This depolarization mechanism does \textit{not} apply to single carriers for which spin-valley locking due to the large spin-orbit spin splittings is expected to lead to significantly \textit{longer} polarization lifetimes.
In the presence of resident carriers, optical excitation
can lead to the formation of charged excitons also called trions, Sec. \ref{sec:charge}. Commonly observed bright trions decay on slightly longer timescales than excitons, namely in about 30~ps at $T=4$~K \cite{Wang:2014b}, which means that the time range for valley index manipulation is still restricted to ultra-fast optics. For future valleytronics experiments and devices, it is therefore interesting to know whether the \textit{resident} carriers left behind after recombination are spin and valley polarized.\\
\indent Several recent time-resolved studies point to encouragingly long polarization dynamics of resident carriers in monolayer TMDs at low temperature. Polarization decays of 3--5 ns were observed in CVD-grown MoS$_2$ and WS$_2$ monolayers that were unintentionally electron-doped \cite{yang:2015a,yang:2015b,Bushong:2016a}, as can be seen in Fig.~\ref{fig:fig7}d.  Longer times up to tens of ns were observed in unintentionally hole-doped CVD-grown WSe$_2$ \cite{hsu:2015,Song:2016a}. Using time-resolved Kerr rotation, the spin/valley dynamics of resident electrons and holes in charge-tunable WSe$_2$ monolayer were recenty measured by \textcite{Dey:2017a}. In the $n$-type regime, long ($\sim$70 ns) polarization relaxation of electrons were observed and considerably longer ($\sim 2 \mu$s) polarization relaxation of holes were revealed in the $p$-doped regime (see Fig.~\ref{fig:fig7}e), as expected because of the strong spin-valley locking of holes in the valence band of monolayer TMDs. Long hole polarization lifetimes were also suggested by a recent report of microsecond hole polarizations of indirect excitons in WSe$_2$/MoS$_2$ bilayers \cite{Kim:2016b}. 
In this case rapid electron-hole spatial separation following neutral exciton generation leads to long-lived indirect excitons, in which the spatial overlap of the electron and hole is relatively small. If the two layers are not aligned with respect to the in-plane angle, there is also an additional mismatch of the respective band extrema in momentum space \cite{Yu:2015z}. The resulting oscillator strength is very small and should directly lead to a rather slow spin-valley depolarization through long-range exchange coupling, previously discussed in Sec.~\ref{sec:depol}. One of the most important challenges at this early stage is to identify the conditions and mechanisms that promote transfer of the optical generated valley polarization of trions or neutral excitons  to the resident carriers \cite{Dyakonov:2008a,Glazov:2012a}. 

\subsection{Lifting valley degeneracy in external fields}
\label{sec:bfield}

\begin{figure}
\includegraphics[width=0.48\textwidth]{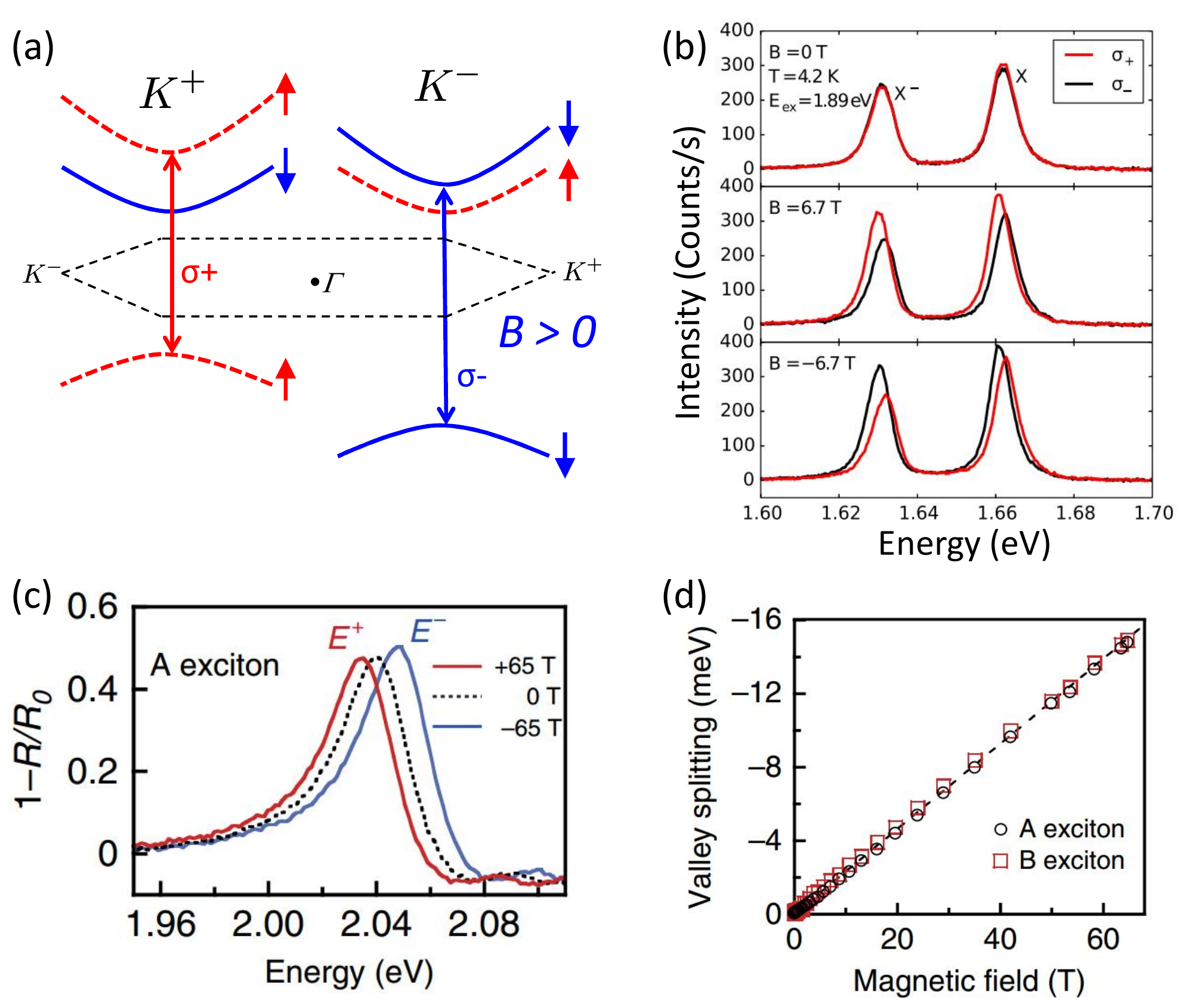}
\caption{\label{fig:fig8} (a) Schematic of Zeeman shifts in magnetic field $\bm B$ perpendicular to the monolayer plane. (b) Measurements on \textbf{MoSe$_2$} MLs from \cite{Macneill:2015a} that show a clear Zeeman splitting. (c) Reflectivity measurements on \textbf{WS$_2$} MLs in high magnetic fields and (d) the Zeeman splitting extracted for A- and B-excitons \cite{Stier:2016a}.
}
\end{figure}

In the absence of any external or effective magnetic or electric field, the exciton transitions involving carriers in the $K^+$ and $K^-$ valley are degenerate and the spin states in the two types of valleys are related by time reversal symmetry. 
This symmetry can be broken through the application of an external magnetic field \emph{perpendicular} to the plane of the monolayer. 
There are two important consequences that are briefly discussed below: first, the valley states  split by a Zeeman energy $\Delta E_Z$ typically on the order of a few meV at 10 Tesla. Second, the valley polarization could change due to this splitting, as the lower energy valley might be populated preferentially. \\
\indent Application of a magnetic field $B_z$ along the $z$ direction (perpendicular to the ML plane) gives rise to a valley Zeeman splitting in monolayer WSe$_2$ and MoSe$_2$ \cite{Li:2014a,Macneill:2015a,Aivazian:2015a,srivastava:2015,Wang:2015d}, lifting the valley degeneracy. In these studies, an energy difference $\Delta E_Z$ on the meV scale is found between the $\sigma^+$ and $\sigma^-$ polarized PL components, stemming from the $K^+$ ($K^-$) valley, respectively, as $\Delta E_Z=E(\sigma^+)- E(\sigma^-)$. In monolayer MoSe$_2$, the $\sigma^+$ and $\sigma^-$  PL components are clearly split in magnetic fields of 6.7~T as shown in Fig.~\ref{fig:fig8} \cite{Macneill:2015a}. The valley Zeeman splitting scales linearly with the magnetic field as depicted in Fig.~\ref{fig:fig8} and the slope gives the effective exciton $g$-factor as $\Delta E_Z=g_X \mu_B B_z$, where $\mu_B$ is the Bohr magneton. The exciton $g$-factor $g_X$ measured for instance in PL  contains contribution from electron and hole $g$-factors. 
In several magneto-optics experiments also on ML MoTe$_2$ and WS$_2$ \cite{Arora:2016a,Stier:2016a,mitioglu:2015, mitioglu:2016} the exciton $g$-factor is about $-4$. The exact energy separation of the valley and spin states is important for spin and valley manipulation schemes. In addition, the $g$-factor also contains important information on the impact of remote bands on the optical transitions, in a similar way as the effective mass tensor, see discussions in \textcite{Wang:2015d,Macneill:2015a}.
The origin of this large $g$-factor is currently not fully understood. There are basically two approaches to calculate the Zeeman splittings in TMD MLs. One is based on the atomic approach by considering atoms as essentially isolated and associating the $g$-factors of the conduction and valence band states with the spin and orbital contributions of corresponding $d_{z^2}$ and $d_{(x\pm i y)^2}$ atomic shells \cite{Aivazian:2015a,srivastava:2015}. The other approach is based on the Bloch theorem and $\bm k \cdot \bm p$-perturbation theory which allows to relate $g$-factor with the band structure parameters of TMD ML \cite{Macneill:2015a}. Merging these approaches, which can be naturally done within atomistic tight-binding models \cite{ Wang:2015d,Rybkovskiy:2017a}, is one of the open challenges for further theoretical studies.\\
\indent At zero magnetic field, the valley polarization in optical experiments is only induced by the circularly polarized excitation laser. At finite magnetic fields, a valley Zeeman splitting is induced and the observed polarization may now also depend on the magnetic field strength. For ML WSe$_2$, sign and amplitude of the valley polarization, even in magnetic fields of several Tesla, is mainly determined by the excitation laser helicity \cite{Wang:2015d,mitioglu:2015}. In contrast, the sign and amplitude of the valley polarization detected via PL emission in MoSe$_2$ and MoTe$_2$ is mainly determined by the sign and amplitude of the applied magnetic field \cite{Wang:2015d,Macneill:2015a,Arora:2016a}.\\
\indent In contrast to a \emph{perpendicular} magnetic field, in monolayer MoS$_2$ an \emph{in-plane} magnetic field ($xy$) up to 9~T does not measurably affect the exciton valley polarization or splitting \cite{Sallen:2012a,Zeng:2012a}, as expected from symmetry arguments. The in-plane field, however, mixed spin-up and spin-down states in the conduction and valence bands and activates spin-forbidden excitons as discussed above in Sec.~\ref{light:dark} and in Refs.~\cite{Zhang:2017a,Molas:2017a}.\\
\indent An elegant, alternative way of lifting valley degeneracy is using the optical Stark effect. Typically a circularly polarized pulsed laser with below bandgap radiation is used to induce a shift in energy of the exciton resonance \cite{Joffre:1989a,Press:2008a}. This shift becomes valley selective in ML TMDs, with induced effective Zeeman splitting is up to $\approx$20 meV, corresponding to effective magnetic fields of tens of Tesla \cite{Sie:2015b,Kim:2014z}. The effective magnetic field created by the Stark effect can also be employed to rotate a coherent superposition of valley states \cite{Ye:2017a}.

\section{Summary and Perspectives}
\label{sec:futur}
In this short review we have detailed some of the remarkable optical properties of transition metal dichalcogenide monolayers. The strong Coulomb interaction leads to exciton binding energies of several hundred meV and excitons therefore dominate the optical properties up to room temperature. The ultimate thinness of these materials provides unique opportunities for engineering the excitonic properties. For example, the dielectric environment can be tuned. Here first experiments show that encapsulation of TMD monolayers in hexagonal boron nitride, for example, significantly reduces the exciton binding energy \cite{Stier:2016a}. More experiments will show in the future how sensitive the exciton ground, excited states and the free carrier bandgap are to changes in their dielectric environment \cite{Ye:2014a,Raja:2017a}, which will depend on the spatial extent of the different states. \\
\indent Another route to engineering the optical properties, particularly the polarization dynamics, is to place ferromagnetic layers close to the monolayer. These proximity effects might be able to lift valley degeneracy even without applying any external magnetic fields, a great prospect for controlling spin and valley dynamics \cite{Zhong:2017a,Zhao:2016a}.\\
\indent In this article we have concentrated on excitons in single monolayers, but many of these concepts apply also to more complex exciton configurations in van der Waals heterostructures \cite{Geim:2013a} where the electrons and holes do not necessarily reside in the same layer. Here many possibilities can be explored, such as studies of Bose-Einstein condensates and superfluidity;  the wide choice of layered materials allows tuning the oscillator strength of the optical transitions as well as the spin- and valley polarization lifetimes \cite{Kim:2016b,Rivera:2016a,Fogler:2014a,Ceballos:2014a,Nagler:2017a}. 

\section*{Acknowledgments}

G.W. and B.U. acknowledge funding from ERC Grant No. 306719.
A.C. gratefully acknowledges funding by the Deutsche Forschungsgemeinschaft (DFG)  through the Emmy
Noether Programme (CH 1672/1-1). 
M.M.G. acknowledges support through RF President grant MD-1555.2017.2 and RFBR projects 17-02-00383 and 17-52-16020. 
T.F.H. wishes to acknowledge support through the AMOS program within the Chemical Sciences, Geosciences, and Biosciences Division, Office of Basic Energy Sciences of the U.S. Department of Energy under Contract No. DE-AC02-76-SFO0515 and by the Gordon and Betty Moore Foundation’s EPiQS Initiative through Grant No. GBMF4545.  
X.M. and T.A. thank ANR MoS2ValleyControl. X.M. also acknowledges the Institut Universitaire de France. M.G, X. M. B.U and T.A. acknowledge the LIA CNRS-Ioffe ILNACS.\\
\indent We thank group members and colleagues, past and present, for stimulating discussions, in particular T. Berkelbach, D. Reichman, C. Robert, I.C. Gerber, M.V. Durnev, M.A. Semina and E.L. Ivchenko.  

\textit{*present address G.W.: Cambridge Graphene Centre, University of Cambridge, Cambridge CB3 0FA, UK}

\bibliographystyle{apsrmp}




\end{document}